\documentclass[sigconf, nonacm, screen]{acmart}

\usepackage{pvldb}

\makeatletter
\@ACM@balancefalse
\makeatother

\AtBeginDocument{%
  }
\renewcommand{\abstractname}{ABSTRACT}

\usepackage{booktabs}
\usepackage{graphicx}
\usepackage{enumitem}
\usepackage{subcaption}
\usepackage{cleveref}
\usepackage{placeins}
\usepackage{xspace}
\usepackage{tikz}
\usetikzlibrary{arrows.meta}
\usepackage[ruled,vlined,linesnumbered]{algorithm2e}
\definecolor{chronoslegendpurple}{HTML}{9467BD}
\definecolor{chronoslegendblue}{HTML}{1F77B4}
\definecolor{chronoslegendorange}{HTML}{FF7F0E}
\definecolor{chronoslegendred}{HTML}{D62728}
\definecolor{chronoslegendgreen}{HTML}{16A34A}
\definecolor{chronoslegendblack}{HTML}{081D15}
\definecolor{swarmnativeorange}{HTML}{D95F02}
\definecolor{swarmlockblue}{HTML}{4C78A8}
\crefname{algocf}{Algorithm}{Algorithms}
\Crefname{algocf}{Algorithm}{Algorithms}

\newcommand{\chronos}{\textsc{Chronos}\xspace}
\newcommand{\chronosfs}{\textsc{ChronosFS}\xspace}
\newcommand{\parhead}[1]{\par\noindent\textbf{#1.}\quad}
\newcommand{\txnbegin}[1]{\textbf{BEGIN} #1}
\newcommand{\txncommit}{\textbf{COMMIT}}

\author{Xinjing Zhou}
\orcid{0009-0008-5949-6553}
\affiliation{%
  \institution{MIT CSAIL}
  \city{Cambridge}
  \state{MA}
  \country{USA}
  }

\author{Jason Mohoney}
\orcid{0000-0001-5497-0481}
\affiliation{%
  \institution{MIT CSAIL}
  \city{Cambridge}
  \state{MA}
  \country{USA}
  }

\author{Samuel Madden}
\orcid{0000-0002-7470-3265}
\affiliation{
  \institution{MIT CSAIL}
  \city{Cambridge}
  \state{MA}
  \country{USA}
}

\author{Michael Stonebraker}
\orcid{0000-0001-9184-9058}
\affiliation{
  \institution{MIT CSAIL}
  \city{Cambridge}
  \state{MA}
  \country{USA}
}

\author{Lei Cao}
\orcid{0000-0001-9909-8607}
\affiliation{
  \institution{University of Arizona}
  \city{Tucson}
  \state{AZ}
  \country{USA}
}

\begin{document}
\raggedbottom

\title{\chronos: Efficient Bolt-On Branching Across Data Stores for Stateful Agentic Applications}

\begin{abstract}

Data-centric applications increasingly use speculative execution to explore multiple candidate paths where each path modifies state distributed across heterogeneous data stores. This trend is intensified by the rise of tool-calling agents. Hence, applications need data systems that can create branches quickly, isolate state-modifying paths, and merge changes consistently across stores without imposing substantial query overhead.
Existing systems provide only partial support, forcing applications to coordinate branches and merges manually, which increases overhead and risks inconsistent cross-store state.

To solve this problem, we introduce \chronos, a bolt-on system that provides branching capability across heterogeneous data stores. We make two contributions. First, \chronos introduces a compact interval-based versioning technique that enables efficient branching and data sharing through simple query rewrite. Second, \chronos introduces a bolt-on architecture that separates branch management from data path within each store. Combined with interval-based versioning, this separation provides atomic cross-store visibility for merges and enables \chronos to support diverse data stores without modifying their engines.

We implement \chronos for PostgreSQL, SQLite, DuckDB, Qdrant, and a DBMS-backed filesystem. We evaluate it using cross-store agent workflows, MCTS-style exploration, and per-store benchmarks. \chronos runs MCTS-style exploration up to 16.7$\times$ faster than existing approaches while maintaining practical query performance across the underlying stores. Under concurrent cross-store workflows, \chronos prevents partially visible merges while substantially outperforming serialized execution.

\end{abstract}

\maketitle

\section{INTRODUCTION}
\label{sec:intro}
\textbf{Versioning for Exploratory Workloads.}
Modern data-centric applications increasingly create isolated versions of durable state to evaluate
tentative changes.
Developers branch databases to test application changes without affecting production
state~\cite{neon-preview-environments,databricks-evolutionary-db-branching,
dolt-use-cases,dolt-database-branches}. Data analysts similarly branch evolving shared datasets
to isolate transformations and preserve their history~\cite{datahub-cidr,datahub-pvldb,
dataset-versioning-principles,decibel,rstore,scientific-array-versioning,
time-travel-arraydb,orpheusdb,tardisdb}. The rapid rise of tool-calling LLM agents both intensifies these established versioning workloads and introduces new
exploratory applications. By automating multi-step development and analysis tasks, agents increase
the frequency and concurrency with which existing workflows create and modify  versions.
At the same time, they  generate and evaluate multiple execution paths at
test time~\cite{llm-test-time-compute,best-of-n-self-certainty,liu2025supporting}.
Through tool calls, each path may update files,
database records, and generated artifacts before evaluation determines which changes to retain or roll back.
This form of speculative execution appears in automated algorithm discovery and machine-learning
optimization~\cite{alphaevolve,openevolve,karpathy-autoresearch,auto-research-survey,
adrs-project,aide}, scientific simulation~\cite{quanscient-agentic-simulation,
aws-simulation-assistant,agentic-scientific-simulation}, and hardware and computational
design~\cite{agentic-hpc-design-exploration,agentic-materials-computation,
agentic-hardware-design-verification}.
Reinforcement learning for tool-using agents introduces another use case. Parallel rollouts
read and modify isolated databases and filesystems before policy updates~\cite{agent-world-model,deepseekmath}.
Each rollout must fork its state quickly, modify it, and run queries that compute its reward.

\begin{figure}[t]
  \centering
  \includegraphics[width=\columnwidth]{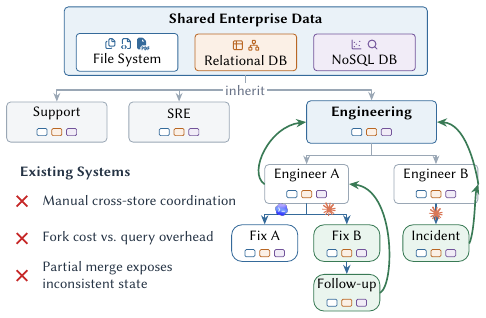}
  \Description{An enterprise data hierarchy rooted at company-wide data and branching into
  department, team, and employee data. An agent creates an isolated task branch spanning files,
  relational records, and vector-search data, then merges accepted changes.}
  \caption{An enterprise agentic platform in which branches inherit shared enterprise data spanning
  filesystem, relational database, and NoSQL database. Existing systems require manual
  cross-store coordination, trade fork cost against query overhead, and can expose inconsistent
  state during a partially visible merge.}
  \label{fig:enterprise-memory}
\end{figure}
\textbf{Cross-Store Speculative State.}
A key problem is that maintaining an isolated version becomes difficult when  logical application state
spans several
data stores. \Cref{fig:enterprise-memory} illustrates this problem in an enterprise agentic
platform
whose shared enterprise data includes source code and documentation, structured records,
and a vector index for semantic retrieval. Support, SRE, and Engineering teams inherit this data
while maintaining additional team data. Within the Engineering team, each employee may use coding
agents that can evaluate competing fixes on private branches, extend a promising fix with a nested branch, and
investigate an unrelated incident concurrently. Each investigation may update source files,
provenance records, and the corresponding vector index. When a reviewed result is merged, these
dependent changes would ideally become visible consistently across all stores so that an entry in vector db never refers to a document that is not yet visible on filesystem. Applications therefore need one branch abstraction
that spans multiple heterogeneous stores, supports fast and flexible branching, efficient query, preserves isolated versions across many tool
calls, and allows changes to be merged atomically or discarded.

\textbf{Existing systems fall short.}
Existing mechanisms cover only parts of these cross-store versioning requirements. Git~\cite{git}
versions source files but does not isolate the associated database state. Database transactions~\cite{gray81,bernstein09}
can isolate changes, but the isolated state is transient because its lifetime is
tied to a single transaction. Physical cloning incurs significant storage and creation costs. Dataset versioning systems and branchable
databases provide durable alternatives~\cite{orpheusdb,tardisdb,dolt,doltgres,neonbranching}, but
can introduce significant query overhead~\cite{ang2026branchbench}. Combining separate mechanisms
for each store leaves the application responsible for aligning their versions and coordinating
every cross-store merge, increasing the complexity of the application and the risk of exposing inconsistent state across stores.

\textbf{\chronos.}
In light of these gaps, we introduce \chronos, a lightweight polystore versioning layer that
provides a unified branch abstraction for application state distributed across heterogeneous stores. An application can fork a
consistent state, let each execution modify its private branch, and then compare, retain, merge, or
discard the result without cloning every store. \chronos operates through existing data store interfaces
and requires no changes to engine internals, while preserving each store's native data and query
paths. Its mechanisms apply independently of
the application, while its design targets frequent, long-running branches such as those created by
stateful agents. Providing frequent branching across existing stores—without modifying their engines or sacrificing query performance—raises three systems challenges. 

\textbf{Challenge \#1: Branching Performance.}
Speculative execution frequently creates branches and executes ordinary queries within them, placing both operations on the critical path. The challenge is to make branch creation independent of the amount of existing data without introducing branch-history reconstruction into the query path. Query execution performance should remain close to the native store as the branch tree grows.

\textbf{Challenge \#2: Store Heterogeneity.}
Speculative execution state may span stores with different data models, query interfaces, and transactional
mechanisms. These stores expose no common branching facility, while a bolt-on layer cannot rely on
changes to their internal implementations. Nevertheless, a branch must select the corresponding
record versions in every participating store and provide the same isolation semantics across them.
The challenge is to enforce this common branch view using only capabilities that heterogeneous stores
already support efficiently, without modifying their engines or replacing their native data and query
paths.

\textbf{Challenge \#3: Cross-store Consistency.}
An accepted candidate may contain dependent changes to files, relational records, and search indexes.
Applying these changes through independent store operations can leave the target in an inconsistent
state if an operation fails or the target changes concurrently. A merge must therefore provide
atomic visibility for the complete resolved result across participating stores, while changes that are not
part of the accepted result remain isolated from the target.

\chronos addresses these challenges with a lightweight interval-based versioning approach and centralized transactional branch coordination. Together, they enable efficient query execution and lightweight branching across heterogeneous stores.

In interval-based record versioning, each branch is assigned a half-open interval, with the root branch spanning the entire interval space. Creating a subbranch partitions its parent's interval, and recursively partitioning intervals naturally represents the branch hierarchy. Each physical record version is tagged with the interval of the branch in which it is created. A record version is visible to a branch when its visibility interval contains the branch's read point, which can be evaluated using a simple predicate. Because a subbranch is assigned an interval contained within its parent's interval, it automatically shares and can read records created by its parent. When a branch updates a shared record, it performs record-level copy-on-write, creating physical versions with disjoint intervals to preserve isolation among branches. This interval-based representation enables efficient query processing using simple filter evaluation without joins.

\chronos addresses the second challenge by applying the interval representation across data
stores, leveraging the filtering capabilities common to most data systems. Branch metadata is
managed and accessed centrally in a relational data store, giving all stores a consistent representation of the branch
hierarchy. Before accessing a branch, \chronos obtains the branch's
interval metadata from the relational store. This interval metadata is then used to augment queries to data stores with interval predicates in a store-specific shim layer. The underlying stores therefore remain responsible for query execution,
indexing, durability, and physical storage, allowing \chronos to support heterogeneous systems
without modifying their engines.

\chronos addresses the third challenge with a merge-publication protocol that provides atomic
cross-store visibility. A merge first stages its record versions without
making them visible. Only after every participating store has durably staged its changes does
\chronos publish the merge by atomically updating the branch metadata in its transactional
relational store. Readers therefore observe either the state before the merge or the
complete merged state, never a partially published merge.

We implement and evaluate \chronos on PostgreSQL, SQLite, DuckDB, Qdrant, and \chronosfs, a
DBMS-backed filesystem. Across PostgreSQL TPC-C and YCSB-C, DuckDB TPC-H, and Qdrant vector search,
\chronos outperforms the next-best versioning alternative by 3.4--17.9$\times$, while incurring a 1.12--1.75$\times$ slowdown relative to unversioned systems.
\chronos completes all five
BranchBench~\cite{ang2026branchbench} workflows. On the branch-intensive
Monte-Carlo-Tree-Search workflow, it is 6.0$\times$ faster than Doltgres and 16.7$\times$ faster
than PostgreSQL cloning on Btrfs.
\chronos is currently deployed at CMS Computing
Operations~\cite{lugato2026archi} at CERN as a storage layer for managing agent knowledge, data
sandboxing and ingestion staging.

In summary, we make the following contributions:
\begin{itemize}[leftmargin=*]
  \item \textbf{A polystore branching abstraction for managing speculative execution state.}
  We design a unified branch API that lets developers and  agentic applications fork, query, diff, merge, and delete
  isolated states across multiple data stores.

  \item \textbf{Interval-based branch management.}
  We present an efficient interval-based data versioning approach for branch creation, record-level copy-on-write, queries, diff,
  and merge without reconstructing versions through membership joins.

  \item \textbf{An extensive evaluation.}
  We evaluate \chronos using cross-store agent workflows, BranchBench~\cite{ang2026branchbench} workloads,
  and microbenchmarks.

\end{itemize}

\section{CHRONOS ARCHITECTURE}
\label{sec:architecture}
\begin{figure}[t]
  \centering
  \includegraphics[width=\columnwidth]{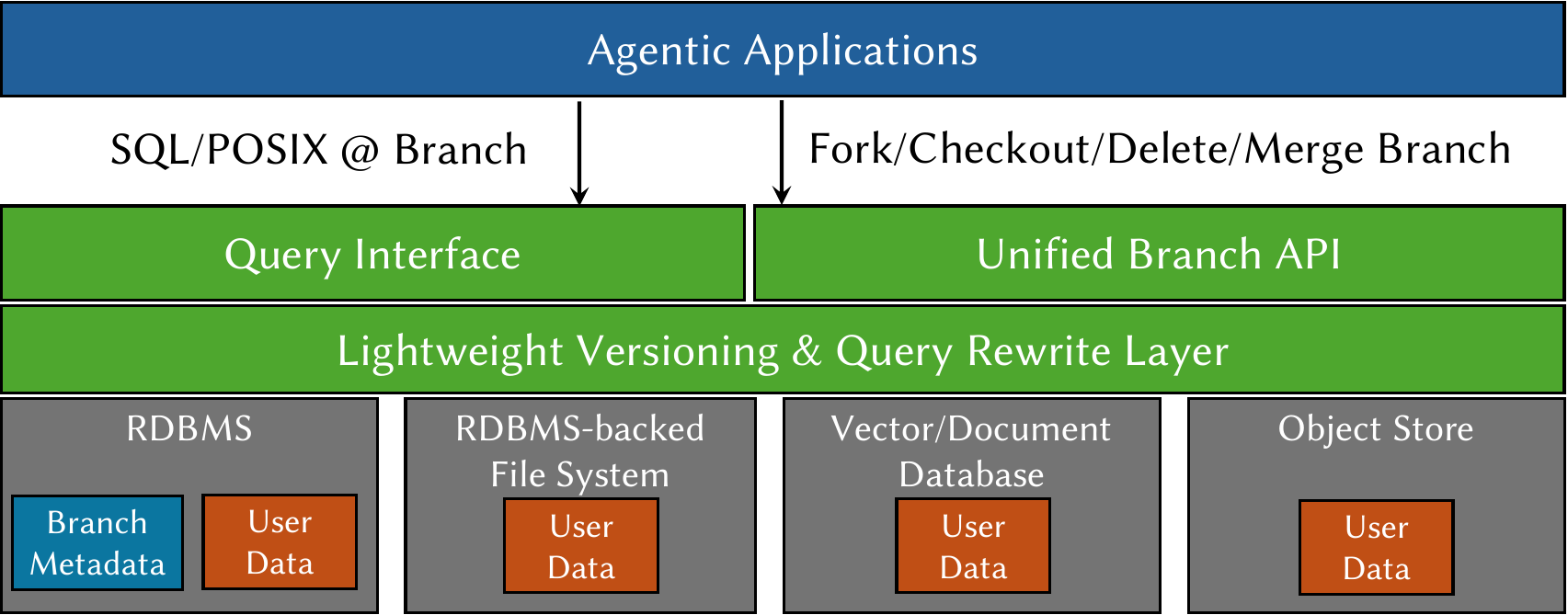}
  \Description{Chronos architecture with an agent application above a Chronos session, a
  transactional metadata store, and store shims for the supported data planes.}
  \caption{\chronos architecture. A transactional metadata store coordinates branch state,
  while store shims connect the branch abstraction to existing data stores.}
  \label{fig:chronos-arch}
  \vspace{-0.5cm}
\end{figure}

In this section, we describe the \chronos architecture, its assumptions, and how applications interact with it.
\Cref{fig:chronos-arch} sketches the architecture. \chronos sits between an agentic application and the data stores that hold branchable state. It uses a
transactional relational database for keeping track of branch metadata and a storage shim for each participating data
store in the \chronos system. The shims are responsible for rewriting queries, maintaining record versions, and acting as a proxy for the application.  Each store handles its own query execution, indexing, durability, and physical storage and is generally unaware of the branching layer.

\subsection{\chronos Assumptions}
\label{sec:assumptions}

\chronos relies on the following assumptions about branch access and data stores.

\begin{itemize}[leftmargin=*]
  \item \textbf{At Least One Transactional Store.} \chronos requires at least one store to support transactions for branch coordination and metadata management.

  \item \textbf{Unique Record Keys.}
  Each record in the branchable state has a stable logical key shared by all of its physical
  versions.

  \item \textbf{Efficient Metadata Filtering.}
  Data stores can keep record metadata and filter it efficiently.

  \item \textbf{Atomic Record Updates.}
  A data store can atomically install the versions produced by one logical write and serialize
  conflicting updates to the same physical versions. Once acknowledged, these versions are durable
  and readable by later \chronos sessions.

  \item \textbf{Controlled Access.} \chronos assume every query sent to each table managed by \chronos is sent through the shim.

\end{itemize}

\subsection{Application Interaction}
\label{sec:semantics}
\label{sec:branching-api}

Applications interact with \chronos through the branch APIs and sessions summarized in
\Cref{tab:branching-api}. A branch identifies one application state across the participating
stores. In the running example shown in \Cref{fig:enterprise-memory}, Engineer A's branch therefore includes source files,
provenance records, and the corresponding search index. To evaluate competing fixes, she uses
\texttt{fork} to create Fix A and Fix B, which initially inherit the same data. She then uses
\texttt{checkout} to create a session for each branch and runs ordinary SQL, search, and filesystem
operations through these sessions. Changes made through a session remain isolated within its branch.
A promising Fix B can be forked again to create the Follow-up branch. After evaluating the candidates, Engineer A uses \texttt{diff} to compare their changes with her
branch. She then uses \texttt{merge} to propagate the accepted Follow-up changes, making the
corresponding file, relational, and search updates visible together. She uses \texttt{delete} to
discard Fix A and reclaim its unreachable versions. The next section explains how interval metadata implements them.

\begin{table}[t]
  \centering
  \caption{Core \chronos branching API}
  \label{tab:branching-api}
  \small
  \begin{tabular}{@{}p{0.27\columnwidth}@{\hspace{0.8em}}p{0.59\columnwidth}@{}}
    \toprule
    \textbf{API} & \textbf{Semantics} \\
    \midrule
    \texttt{fork(source, hint)} &
    Create a child branch from an existing branch with a hint about the branch subtree structure
    (\Cref{sec:create}). \\
    \texttt{checkout(branch)} &
    Create a session for accessing a branch. \\
    \texttt{diff(b1,b2)} &
    Compute changes of a branch with respect to a target branch
    (\Cref{sec:diff-merge}). \\
    \texttt{merge(src,dst)} &
    Merge changes with atomic cross-store visibility
    (\Cref{sec:diff-merge}). \\
    \texttt{delete(branch)} &
    Remove a branch subtree and reclaim its physical versions
    (\Cref{sec:delete-gc}). \\
    \bottomrule
  \end{tabular}
\end{table}

\begin{figure}[t]
  \centering
  \vspace{-4pt}
  \includegraphics[width=\columnwidth, trim=0 6pt 0 14pt, clip]{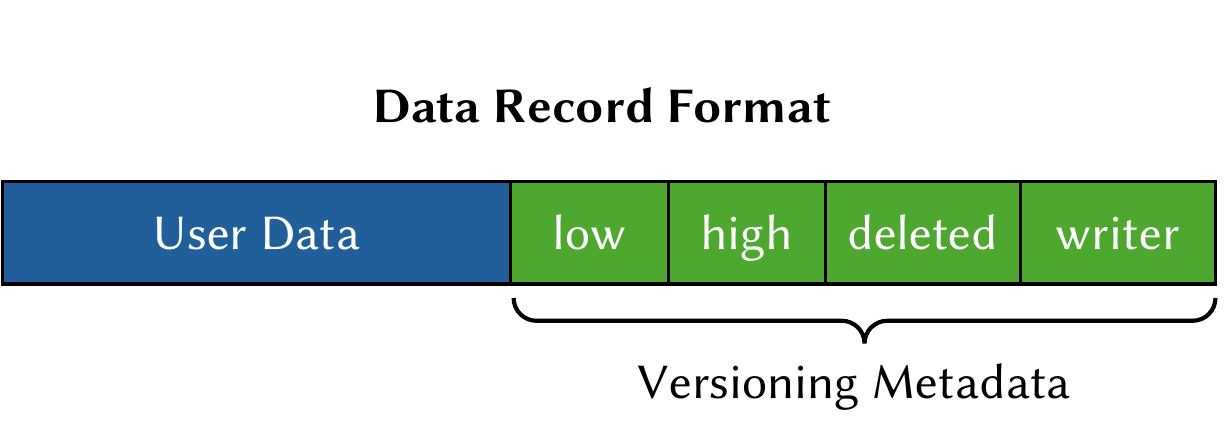}
  \par\vspace{-5pt}
  \Description{A Chronos physical record consists of user data followed by interval low and high
  endpoints, a deletion flag, and a writer identifier. The latter fields form the versioning
  metadata.}
  \caption{\chronos record format. Each physical version stores application data together with
  interval visibility metadata, a writer identifier, and deletion state.}
  \label{fig:record-format}
  \vspace{-0.5cm}
\end{figure}

\section{INTERVAL-BASED DATA ISOLATION AND RECORD UPDATES}
\label{sec:algorithm}
In this section, we describe the interval-based versioning model that underlies \chronos.
The model compactly represents both the branch hierarchy and versioned application state, enabling
efficient branch operations and query execution. \chronos uses integer intervals as lightweight
metadata to encode which record versions are visible to each branch. It rewrites queries with simple
predicates over this metadata, allowing each branch to access its visible state without traversing
branch history. The same representation supports record-level copy-on-write, allowing branches to
share unchanged data while isolating their modifications.
In this section, we describe the high-level versioning scheme and its invariants, and then, in the next section, provide details about how intervals are allocated and how specific operations on branches are implemented.
For clarity,  we assume the data store is a relational database in this section. The same technique applies to other data models as long as records can carry interval metadata and the store can efficiently filter on that metadata.
\subsection{Version Representation and Reads}
\label{sec:model}
\label{sec:reads}

\chronos represents the branch hierarchy using an integer range. Each branch owns a half-open
interval \([l,h)\), and each child receives a subinterval of its parent's interval. Interval
containment therefore encodes branch ancestry without storing it with each
record. Each branch also maintains an allocation frontier \(f\), initially \(l+1\). The frontier
advances monotonically as the branch creates children, partitioning its interval into ranges
allocated to earlier branches and an active range \([f,h)\) for future writes and forks.

\chronos uses the same interval representation to encode record visibility.
\Cref{fig:record-format} shows the physical record layout. Each physical version of a logical
record is tagged with a visibility interval \([\texttt{low},\texttt{high})\) in addition to its application
data. The \texttt{deleted} bit represents a tombstone. The \texttt{writer} field identifies the active range in which a version was created. Each active range has a unique 4-byte writer identifier stored in every version created there.

For reads, the branch frontier \(f\) acts as a read point. A physical version is visible when its
interval contains \(f\) and not a tombstone:
\[
  \texttt{low} \leq f < \texttt{high}
  \quad \wedge \quad
  \texttt{deleted} = \texttt{false}
\]
Thus, \chronos can select the visible version using an ordinary range predicate rather than traversing branch ancestry.

\begin{figure*}[t]
  \centering
  \makebox[\textwidth][c]{%
  \begin{minipage}{1.1\textwidth}
  \centering
  \begin{subfigure}[t]{0.245\linewidth}
    \centering
    \includegraphics[width=\linewidth,trim=14pt 11pt 14pt 11pt,clip]{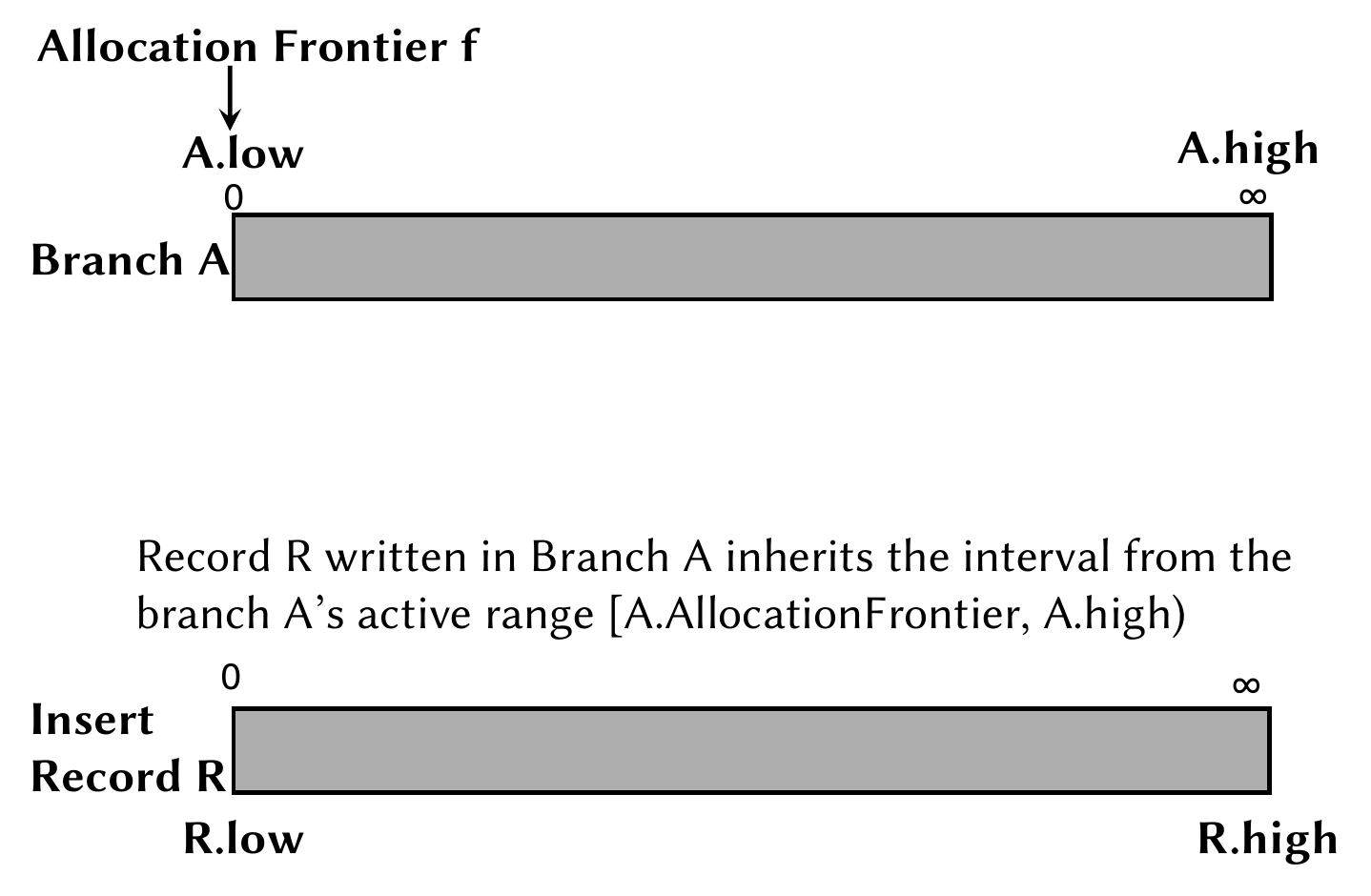}
    \caption{Initial state}
    \label{fig:interval-initial}
  \end{subfigure}\hfill%
  \begin{subfigure}[t]{0.245\linewidth}
    \centering
    \includegraphics[width=\linewidth,trim=14pt 11pt 14pt 11pt,clip]{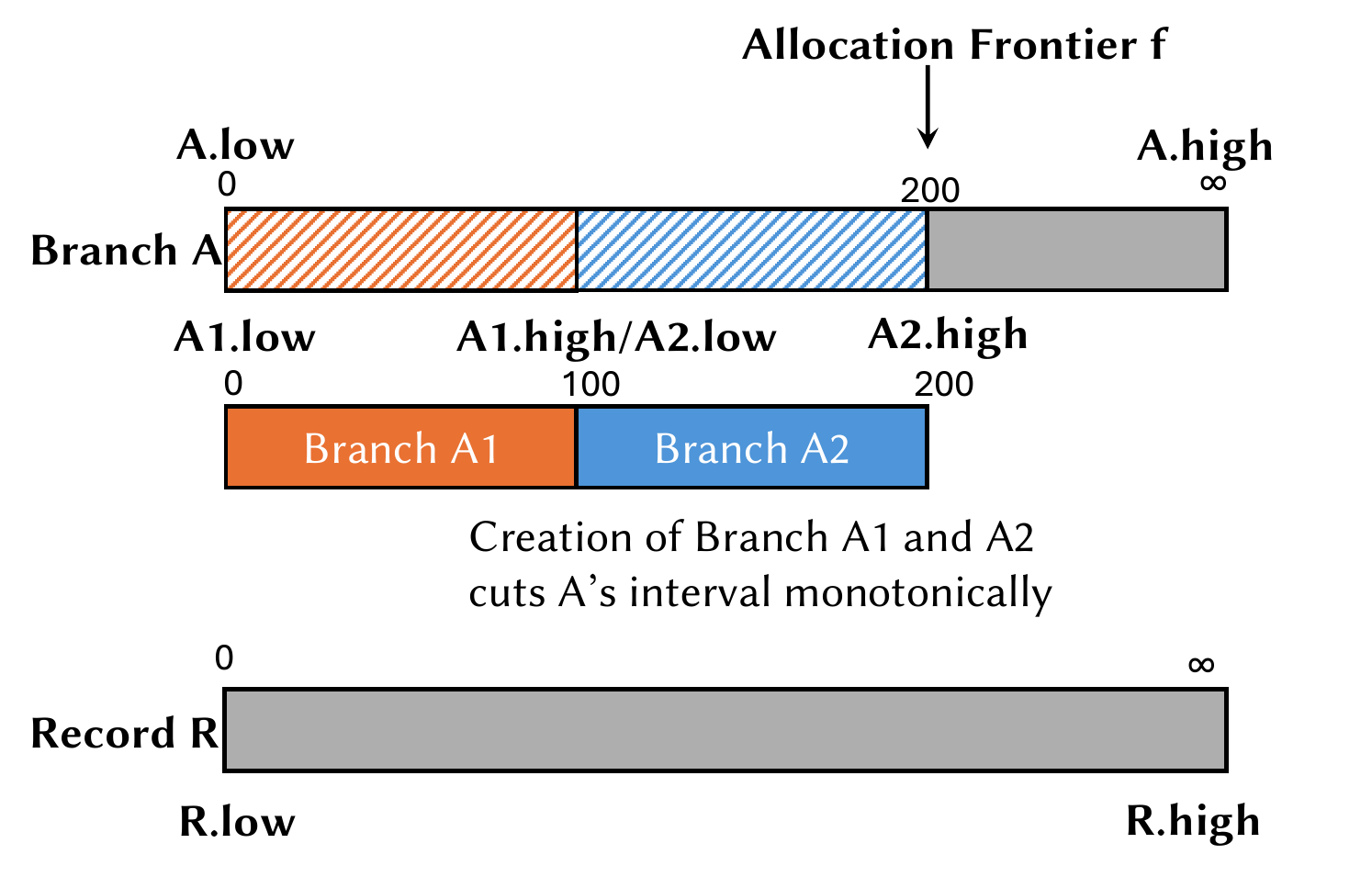}
    \caption{Create branches}
    \label{fig:interval-branch-creation}
  \end{subfigure}
  \hfill
  \begin{subfigure}[t]{0.245\linewidth}
    \centering
    \includegraphics[width=\linewidth,trim=14pt 11pt 14pt 11pt,clip]{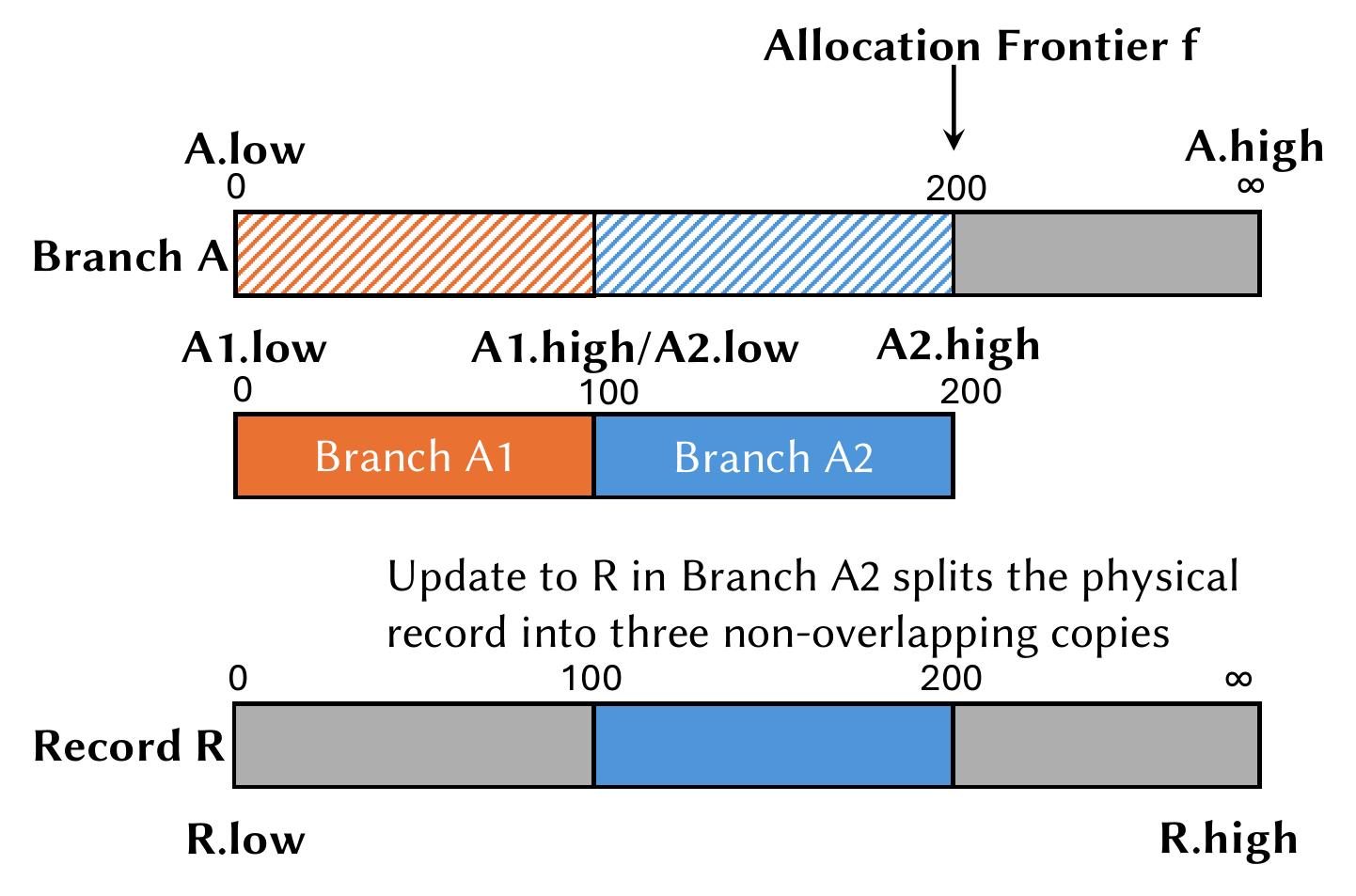}
    \caption{Update record}
    \label{fig:interval-branch-record-update}
  \end{subfigure}
  \hfill
  \begin{subfigure}[t]{0.245\linewidth}
    \centering
    \includegraphics[width=\linewidth,trim=14pt 11pt 14pt 11pt,clip]{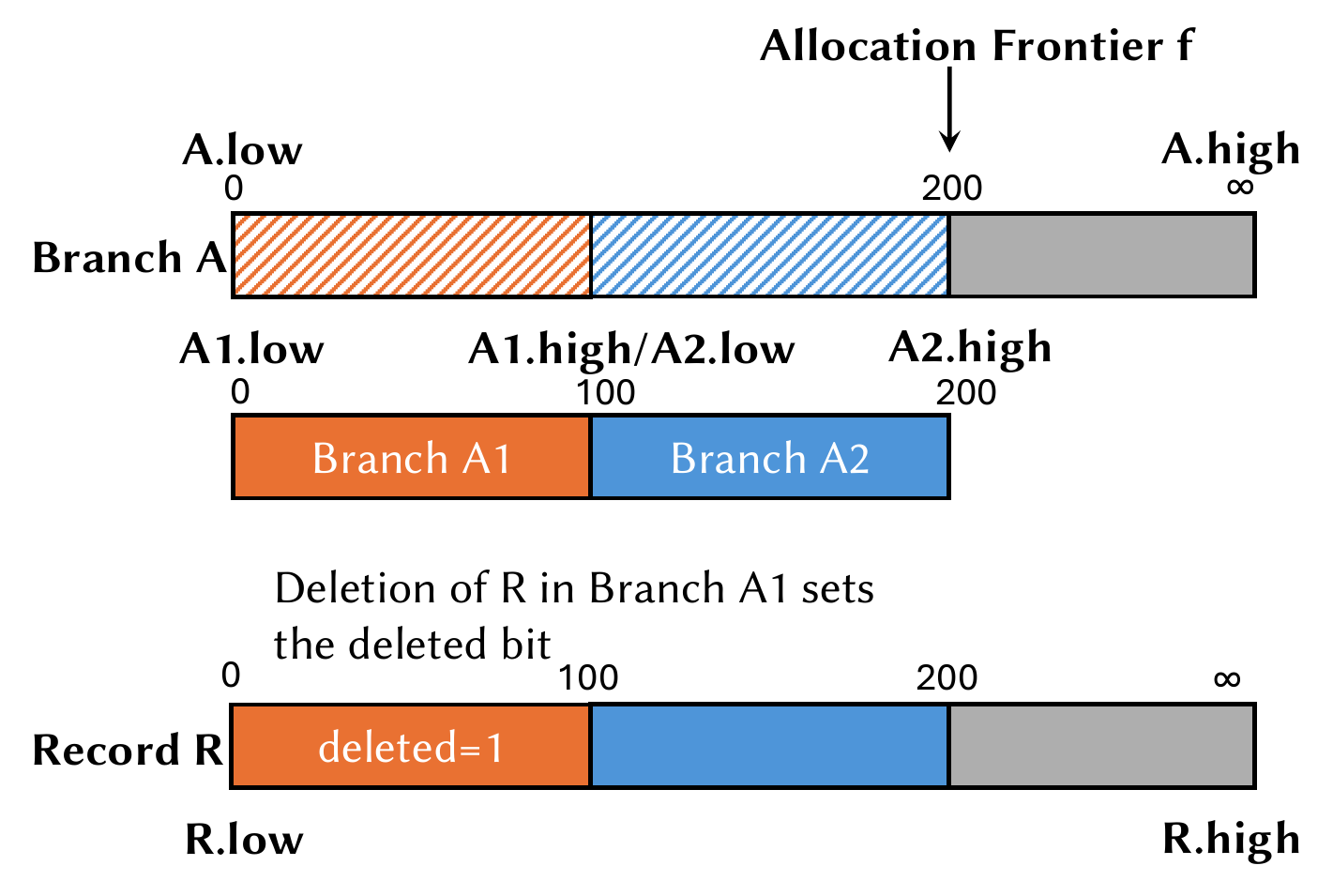}
    \caption{Delete record}
    \label{fig:interval-branch-record-delete}
  \end{subfigure}
  \end{minipage}}
  \Description{Four stages of interval branching. A root branch initially sees a record across its
  interval. Creating two children reserves disjoint subintervals without copying the record. An
  update replaces visibility over one child's interval, and a deletion creates a tombstone over the
  other child's interval.}
  \caption{Overview of \chronos interval operations. Initially, record R spans branch A's
  interval. Two forks reserve disjoint subintervals for branches A1 and A2 without copying record R.
  An update in branch A2 replaces only branch A2's range, while a deletion in branch A1 installs a
  tombstone tagged with branch A1's interval. }
  \label{fig:interval-branching-example} 
\end{figure*}

\Cref{fig:interval-branching-example} illustrates this mechanism. Initially, record R's visibility interval spans branch A's active range, so A can read R. Creating branches A1 and A2 assigns them disjoint subintervals without modifying R. Because R's visibility interval also contains the read points of A1 and A2, all three branches share the same physical version. Forking therefore copies no record data; a new physical version is created only when a branch modifies R.

\parhead{Non-overlapping interval invariant}
For each logical record, \chronos ensures the visibility intervals of any two distinct physical versions \(v_i\)
and \(v_j\) do not overlap:
\begin{equation}
[\texttt{low}_i,\texttt{high}_i) \cap
[\texttt{low}_j,\texttt{high}_j) = \varnothing .
\label{eq:interval-invariant}
\end{equation}
The invariant ensures that the visibility predicate selects at most one physical version of each
logical record.

\parhead{Logical query rewrite}
Let $R$ be a logical relation with application attributes $A_R$, and let $\widehat{R}$ be its physical
relation containing the application attributes and the interval metadata. At frontier $f$ of a branch, the logical
contents of $R$ are
\begin{equation}
R_f = \pi_{A_R}\!\left(
  \sigma_{\texttt{low} \leq f < \texttt{high}\;\wedge\;\neg\texttt{deleted}}
  (\widehat{R})
\right).
\label{eq:visible-relation}
\end{equation}
The selection filters out physical versions not visible to the branch, and the projection removes the interval
metadata. By \Cref{eq:interval-invariant}, $R_f$ contains at most one physical version of
each logical record and is therefore an ordinary relation representing $R$ at frontier $f$. To evaluate a query at frontier $f$ of the branch, \chronos applies a rewrite using to every
 scan in the logical plan over an interval-versioned relation. The scan reads the physical relation
$\widehat{R}$ under the visibility predicate and exposes its application attributes as $R_f$. This rewrite is compositional because it substitutes the branch-visible relation independently at
each table scan. Hence, joins, aggregations, and other operators can evaluate over the resulting
$R_f$ relations, preserving the original logical plan above the scans. Each rewritten scan adds one
fixed predicate whose complexity is independent of branch depth and history.

\subsection{Writes and Interval Splitting}
\label{sec:writes}

To ensure isolation among branches, all writes maintain the non-overlap invariant defined in \Cref{eq:interval-invariant} over the branch's
active range \(w=[f,h)\).
\chronos uses record-level copy-on-writes that never overwrite a physical version's data visible to another branch.
An insert with no overlapping version creates a new physical version tagged with \(w\) and \(w\)'s writer identifier. The branch and its future descendants  share this version.

Updates and deletes are more interesting when the target version is inherited from another branch.
Overwriting the inherited version would change the value observed by other branches. Instead,
\chronos splits its visibility interval at \(w\)'s boundaries. Portions outside \(w\) retain
the inherited value and writer identifier; the portion within \(w\) is replaced by the
new value or a tombstone tagged with \(w\)'s writer identifier. The resulting
intervals remain disjoint.

\Cref{fig:interval-branch-record-update,fig:interval-branch-record-delete} illustrate this operation. In the running example, updating
record R in branch A2 replaces only the interval assigned to A2, leaving the old value visible to branch A and A1.
Deleting record R in branch A1 similarly installs a tombstone only over A1's interval, without affecting the versions visible to branch A or A2. Unmodified ranges therefore remain shared across branches.

\Cref{algo:chronos-split} sketches the interval split procedure for one logical key. It first finds
physical versions whose visibility intervals overlap \(w\), preserves any portions outside \(w\),
and installs a replacement over \(w\). If no version overlaps \(w\), an insert creates the first
version; an update or delete is a no-op. The retained fragments and replacement occupy disjoint
intervals, preserving \Cref{eq:interval-invariant}.

\begin{algorithm}[ht]
\small
\SetKwProg{Fn}{Function}{}{}
\DontPrintSemicolon
\KwIn{operation $op \in \{\textsc{insert},\textsc{update},\textsc{delete}\}$, logical key $key$,
active range $w=[f,h)$, branch writer id $writer$, and $payload$ for insert/update}

\Fn{\textsc{write}(op, key, w, writer, payload)}{
    \eIf{$op=\textsc{delete}$}{
        $rep \gets$ tombstone\;
    }{
        $rep \gets payload$\;
    }
    $V \gets$ physical versions of $key$ that overlap $w$\;
    \If{$V=\varnothing$}{
        \If{$op=\textsc{insert}$}{
            insert a new version with $rep$, interval $w$, and $writer$\;
        }
        \Return\;
    }
    \ForEach{$v \in V$}{
        $R \gets \varnothing$\;
        \If{$v.low < w.low$}{
            $R \gets R \cup \{v\text{ tagged with interval }[v.low,w.low)\}$\;
        }
        \If{$w.high < v.high$}{
            $R \gets R \cup \{v\text{ tagged with interval }[w.high,v.high)\}$\;
        }
        replace $v$ with the versions in $R$\;
    }
    insert a new version with $rep$, interval $w$, and $writer$\;
}

\caption{Copy-on-write Record Interval Split}
\label{algo:chronos-split}
\end{algorithm}

\parhead{Concurrent interval splits}
Sibling branches may update the same inherited physical version even though their active ranges
are disjoint. Their writes can therefore attempt to split the same physical version concurrently.
\chronos serializes such splits using the store's isolation mechanism to atomically install
the retained fragments and replacement versions. For stores that support transactions for isolation (i.e., most databases\cite{postgresql-transactions,mysql-transactions,sqlserver-transactions,sqlite-transactions,
oracle-transactions,mongodb-transactions}),  a transaction can be used to group splits across multiple
keys, so readers observe either the state before the logical write or its complete result.
Stores with weaker transactional support can use other mechanisms, such as atomic
batches, when the affected versions are co-located
\cite{firestore-transactions-batched-writes,cosmosdb-transactional-batch,rocksdb-writebatch,
redis-transactions}.

\parhead{Discussion}
\chronos's interval-based design optimizes for reads and simplicity. A branch read adds only a range predicate, regardless of how deep the branch is. Therefore the cost moves to writes: updating or deleting an inherited version
may create mutliple physical versions, increasing storage and index maintenance overhead. Writes from additional active ranges can thus increase write
amplification and the number of physical versions examined by scans. \chronos trades this write
and storage overhead for a simple, branch-independent read path and metadata-only forks.
This overhead is most pronounced when forks lead several branches to modify the same key. A fork creates
separate writable intervals for the child and the parent. The first write in one of these
intervals can split an inherited version into at most three versions: an unchanged part before the
interval, the replacement within it, and an unchanged part after it. Each written interval
therefore adds at most two live versions. If \(n\) writable intervals created by forks modify the
same key, the key has at most \(2n+1\) live versions. Later updates or deletes in the same interval
reuse the existing split and do not increase this count. A write that overlaps \(s\) existing
versions replaces them with at most three versions and requires \(O(s)\) row changes.

\section{BRANCH OPERATIONS}
\label{sec:branch-operations}

The previous section showed how interval metadata isolates branch-local versions while allowing
unchanged data to remain shared. This section describes how \chronos creates and maintains the
branch metadata that drives this representation, including branch creation, interval allocation,
merge, and deletion. At a high level, \chronos stores this metadata in a transactional relational store so that
branch operations update the hierarchy and interval assignments atomically, which significantly simplifies branch management and coordination.

\subsection{Branch Creation}
\label{sec:create}
\Cref{algo:chronos-create} sketches the metadata changes for branch creation. Branch creation reserves a subinterval from the parent's active range and advances the parent's
frontier in one metadata transaction. Given a parent branch with active range \([f,h)\), the
allocator chooses a split point \(m\). \chronos reserves \([f,f+1)\) for the \emph{fork base}, a
special read-only branch that represents the parent's state at the time of the fork which will be used for merging in \Cref{sec:diff-merge}. It assigns
\([f+1,m)\) to the child with frontier \(f+1\) and advances the parent's frontier to \(m\). The
child and parent thus receive disjoint active ranges for subsequent writes, while both continue to
see the versions visible at the original frontier \(f\). Therefore, branch creation updates only metadata and
copies no record versions.

\begin{algorithm}[th]
\small
\SetKwFunction{FFrontier}{frontier}
\SetKwFunction{FHigh}{high}
\SetKwFunction{FChooseSplit}{choose\_split}
\SetKwFunction{FCreateForkBase}{create\_fork\_base}
\SetKwFunction{FCreateChild}{create\_child}
\SetKwFunction{FAdvance}{advance\_frontier}
\SetKwProg{Fn}{Function}{}{}
\DontPrintSemicolon

\Fn{\textsc{create\_branch}(parent, name, hint)}{
    \txnbegin{metadata transaction}\;
    $f \gets$ \FFrontier{$parent$}, $h \gets$ \FHigh{$parent$}\;
    $m \gets$ \FChooseSplit{$parent.low, f, h, hint$}\;
    \tcp{Reserve $[f,f+1)$ as the read-only fork base}
    \FCreateForkBase{$parent, name, f$}\;
    \tcp{assign $[f+1,m)$ to the child branch}
    \FCreateChild{$name, [f + 1, m), f + 1$}\;
    \FAdvance{$parent, m$}\;
    \txncommit\;
    \Return $name$\;
}

\caption{Branch Creation}
\label{algo:chronos-create}
\end{algorithm}

\subsection{Epoch-Based Coordination}
\label{sec:epoch-coordination}
Branch creation must preserve data isolation for sessions that execute concurrently with interval
reassignment in a branch. Although \Cref{algo:chronos-create} changes the interval metadata atomically,
a session may cache the parent's interval before the fork and issue a later write after part of that
interval has been assigned to the child. Such a stale write could enter the child's range and violate
isolation. More generally, the same anomaly can accompany any change to a branch frontier or writable
interval. Refreshing the cached state before every query would eliminate the race, but would place
synchronization overhead on every query.

To minimize query-processing overhead, \chronos instead uses epoch-based coordination inspired by
epoch-based reclamation~\cite{fraser2004practical}. \chronos associates an
epoch with each branch's interval metadata. A session may cache the interval metadata, but it also
tracks the branch epoch associated with the cached state. Whenever the interval metadata needs to
change, \chronos advances the branch epoch and waits for sessions in the old epoch to finish their
ongoing operations. These sessions periodically refresh their cached interval metadata before entering the
new epoch. The epoch change therefore forms a barrier between operations using the old and new
interval assignments. Crucially, enforcing this barrier does not require per-query metadata refresh. \chronos periodically
tracks which branch epoch each session has observed, independently of query execution. A session can
therefore reuse its cached interval metadata while remaining in the same epoch.
An interval-changing operation waits only for sessions using the old branch state;
otherwise, it proceeds immediately.

\subsection{Interval Allocation}
\label{sec:interval-allocation}
Branch creation leaves one policy question: how should the allocator choose the split point \(m\)?
Giving a child a larger range preserves more capacity for its descendants but leaves less capacity
for additional siblings. The allocator must therefore
balance branch depth and fanout within a fixed-width integer domain. Consider a root-to-leaf path on which the branch at depth \(d\) may
create \(F_d\) children. Preserving an active range for the parent and one for each child requires
approximately \(\lceil\log_2(F_d+1)\rceil\) coordinate bits at that level. A path of depth \(D\)
therefore requires
\begin{equation}
  b \gtrsim
  \sum_{d=0}^{D-1}\left\lceil\log_2(F_d+1)\right\rceil,
  \label{eq:interval-capacity}
\end{equation}
where \(b\) is the coordinate width. This bound exposes the fundamental width--depth tradeoff of
any nested-interval representation. An allocator can distribute the available bits according to
the expected tree shape, but it cannot guarantee unbounded fanout and depth without widening the
coordinates or reassigning existing intervals.

However, branching is rarely uniform across depth in real-world workloads. For example, a study~\cite{zou2019branchuse} shows that among 2,923 open-source projects hosted on GitHub, only 91 (3.1\%) created more than 30
branches. With the rise of agentic applications, we expect the need for branching will increase mostly from search-oriented workloads~\cite{ang2026branchbench}. These workloads may create many more branches, but
their broadest expansion commonly occurs near the root, after which only promising alternatives
are extended. Many branches are also created for one-off use cases such as debugging or simulation. Those branches do not branch further, which we call \emph{terminal} branches.

Based on these insights, \chronos proposes a hybrid policy in \Cref{algo:chronos-allocate} that adapts to both known and unknown fanout scenarios using application hints. When the application knows that a branch will
create \(k\) children, the allocator divides its initial capacity into \(k+1\) equal-width ranges,
one for each child and one retained by the parent. When fanout is unknown, \chronos reserves equal-width child slots near the top of the tree, where
fanout is most likely to consume a large fraction of the available coordinate space. At depth
\(d<3\), reserving \(r_d\) bits provides up to \(2^{r_d}-1\) child slots while retaining one range
for the parent. We choose \(r_0 \ge r_1 \ge r_2\), decreasing the reservation with depth. Outside these reserved slots, \chronos uses an adaptive harmonic allocation when the fanout
is unknown. Let \(R_n\) be the active capacity after \(n\) such children and let \(q\) be a
configurable fanout reserve. The next child receives \(R_n/(q+n+1)\), leaving
\(R_{n+1}=R_n(q+n)/(q+n+1)\) for the parent. Equivalently, relative to the capacity \(R_0\) when
adaptive allocation begins,
\[
  R_n = R_0\frac{q}{q+n}, \qquad
  s_n = R_0\frac{q}{(q+n-1)(q+n)},
\]
where \(s_n\) is the range assigned to the \(n\)-th child. The reserve prevents the first unknown
child from consuming half of the range. The parent's active capacity decreases harmonically with
the observed fanout. We empirically find that \(q=8\) is a good value for most use cases, while allowing applications to configure it for different expected
fanouts. Finally, a terminal branch receives only the minimum writable range.

\begin{algorithm}[ht]
\small
\SetKwProg{Fn}{Function}{}{}
\DontPrintSemicolon

\Fn{\textsc{choose\_split}(low, f, high, hint)}{
    $initial \gets high - low - 1$\;
    $active \gets high - f - 1$\;
    $d \gets hint.depth$\;
    \uIf{$hint.terminal$}{
        $child\_width \gets 1$\;
    }
    \uElseIf{$hint.fanout > 0$}{
        \tcp{Known fanout: use equal ranges}
        $child\_width \gets \lfloor initial/(hint.fanout+1) \rfloor$\;
    }
    \uElseIf{$d < 3$}{
        \tcp{Unknown fanout near root: use reservation $r_d$}
        $child\_width \gets \lfloor initial/2^{r_d} \rfloor$\;
    }
    \Else{
        \tcp{Unknown fanout: adaptive harmonic allocation}
        $n \gets hint.children\_created$\;
        $child\_width \gets \lfloor active/(q+n+1) \rfloor$\;
    }
    \Return $f + 1 + child\_width$\;
}

\caption{Hybrid Integer-Interval Allocation}
\label{algo:chronos-allocate}
\end{algorithm}

\subsection{Atomic Visibility for Cross-Store Merge}
\label{sec:diff-merge}

\begin{figure}[t]
  \centering
  \resizebox{0.98\columnwidth}{!}{\begin{tikzpicture}[
  x=1cm,
  y=1cm,
  font=\scriptsize,
  bar/.style={draw=black, line width=0.7pt},
  prior/.style={bar, fill=swarmnativeorange!28},
  consumed/.style={bar, fill=swarmnativeorange!14},
  active/.style={bar, fill=chronoslegendgreen!24},
  staged/.style={bar, fill=swarmlockblue!16},
  published/.style={bar, fill=swarmlockblue!88},
  frontier/.style={draw=swarmnativeorange!90!black,
    -{Stealth[length=1.55mm,width=1.25mm]}, line width=0.65pt,
    line cap=round, line join=round},
  stagearrow/.style={draw=black,
    -{Stealth[length=1.45mm,width=1.15mm]}, line width=0.55pt,
    line cap=round, line join=round}
]

\node[font=\bfseries] at (4.25,1.32) {(a) Reserve the staging range};
\node at (4.25,0.82)
  {$f_{\mathrm{new}}=f_{\mathrm{old}}+2<h$};

\node[anchor=east, font=\bfseries, align=right] at (0.97,-0.22)
  {Target Branch\\Interval};
\fill[swarmnativeorange!28] (1.10,0.01) rectangle (3.55,-0.45);
\fill[swarmlockblue!16] (3.55,0.01) rectangle (5.20,-0.45);
\fill[chronoslegendgreen!24] (5.20,0.01) rectangle (7.90,-0.45);
\draw[black, line width=0.7pt] (1.10,0.01) rectangle (7.90,-0.45);
\draw[black, line width=0.7pt] (3.55,0.01) -- (3.55,-0.45);
\draw[black, line width=0.7pt] (5.20,0.01) -- (5.20,-0.45);
\node at (4.375,-0.22) {staging range};
\node at (6.55,-0.22) {new active range};

\draw (1.10,0.01) -- (1.10,0.18);
\node[anchor=south] at (1.10,0.19) {$low$};
\draw (2.45,0.01) -- (2.45,0.18);
\node[anchor=south] at (2.45,0.19) {$f_{\mathrm{old}}$};
\draw (3.55,0.01) -- (3.55,0.18);
\node[anchor=south] at (3.55,0.19) {$f_{\mathrm{old}}+1$};
\draw (5.20,0.01) -- (5.20,0.18);
\node[anchor=south] at (5.20,0.19) {$f_{\mathrm{new}}$};
\draw (7.90,0.01) -- (7.90,0.18);
\node[anchor=south] at (7.90,0.19) {$h$};

\draw[frontier] (2.45,-0.87) -- (2.45,-0.47);
\node[anchor=east] at (2.32,-0.67) {frontier};
\node[anchor=north] at (5.05,-0.98)
  {$I_m=[f_{\mathrm{old}}+1,h)$ remains hidden at $f_{\mathrm{old}}$};

\draw[black!35, line width=0.45pt] (0.20,-1.38) -- (8.15,-1.38);

\node[font=\bfseries] at (4.25,-1.72) {(b) Stage merge writes in participating stores};
\node at (4.50,-2.06) {resolved changes $\Delta$};
\draw[stagearrow] (4.28,-2.16) -- (2.125,-2.28);
\draw[stagearrow] (4.50,-2.16) -- (4.505,-2.28);
\draw[stagearrow] (4.72,-2.16) -- (6.875,-2.28);
\path[staged] (1.10,-2.38) rectangle (3.15,-2.86);
\path[staged] (3.48,-2.38) rectangle (5.53,-2.86);
\path[staged] (5.85,-2.38) rectangle (7.90,-2.86);
\node at (2.125,-2.62) {\shortstack{Relational store\\$\Delta$ over $I_m$}};
\node at (4.505,-2.62) {\shortstack{NoSQL database\\$\Delta$ over $I_m$}};
\node at (6.875,-2.62) {\shortstack{Filesystem\\$\Delta$ over $I_m$}};
\node[anchor=north] at (4.50,-2.96)
  {Durably staged in every store; the target branch remains at $f_{\mathrm{old}}$};

\draw[black!35, line width=0.45pt] (0.20,-3.32) -- (8.15,-3.32);

\node[font=\bfseries] at (4.25,-3.66) {(c) Publish with one metadata transaction};

\node[anchor=east, font=\bfseries, align=right] at (0.97,-4.58)
  {Target Branch\\Interval};
\fill[swarmnativeorange!28] (1.10,-4.35) rectangle (3.55,-4.81);
\fill[swarmlockblue!16] (3.55,-4.35) rectangle (5.20,-4.81);
\fill[chronoslegendgreen!24] (5.20,-4.35) rectangle (7.90,-4.81);
\draw[black, line width=0.7pt] (1.10,-4.35) rectangle (7.90,-4.81);
\draw[black, line width=0.7pt] (3.55,-4.35) -- (3.55,-4.81);
\draw[black, line width=0.7pt] (5.20,-4.35) -- (5.20,-4.81);
\node at (4.375,-4.58) {staging range};
\node at (6.55,-4.58) {new active range};

\draw (1.10,-4.35) -- (1.10,-4.18);
\node[anchor=south] at (1.10,-4.17) {$low$};
\draw (2.45,-4.35) -- (2.45,-4.18);
\node[anchor=south] at (2.45,-4.17) {$f_{\mathrm{old}}$};
\draw (3.55,-4.35) -- (3.55,-4.18);
\node[anchor=south] at (3.55,-4.17) {$f_{\mathrm{old}}+1$};
\draw (5.20,-4.35) -- (5.20,-4.18);
\node[anchor=south] at (5.20,-4.17) {$f_{\mathrm{new}}$};
\draw (7.90,-4.35) -- (7.90,-4.18);
\node[anchor=south] at (7.90,-4.17) {$h$};

\draw[frontier] (5.20,-5.23) -- (5.20,-4.83);
\node[anchor=east] at (5.07,-5.03) {frontier};
\node[anchor=north] at (5.65,-5.34)
  {$I_m=[f_{\mathrm{old}}+1,h)$: merge versions become visible at $f_{\mathrm{new}}$};

\end{tikzpicture}}
  \Description{Three stages of cross-store merge publication. Chronos reserves a new interval,
  stages versions carrying that interval in participating stores while the old frontier remains
  visible, and atomically advances the branch frontier to publish all staged versions together.}
  \caption{Cross-store merge publication. (a) \chronos reserves $I_m$ outside old frontier
  $f_{\mathrm{old}}$. (b) Participating stores durably stage their resolved writes over $I_m$ while
  the target branch remains unchanged. (c) One metadata transaction advances the target branch
  frontier to $f_{\mathrm{new}}\in I_m$ and clears its reservation, publishing all writes together.}
  \label{fig:atomic-merge-visibility}
  \vspace{-0.35cm}
\end{figure}
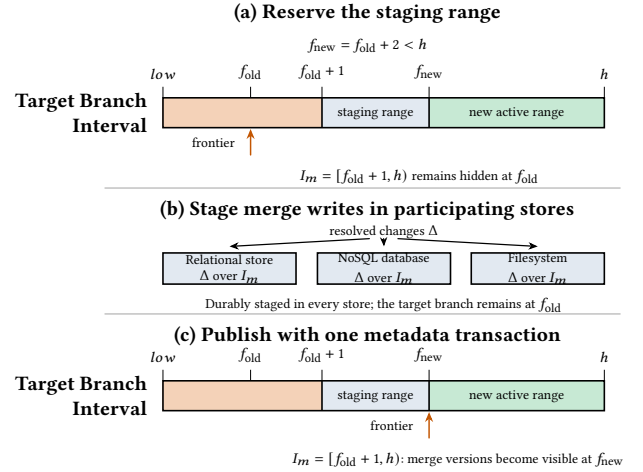

After evaluating a branch, an application may incorporate some or all of its changes into a target
branch. A cross-store merge should move the target branch from one complete logical state to another: a
reader must not observe merged relational data without the corresponding filesystem and vector database
updates. \chronos provides this atomic-visibility guarantee without requiring the participating
stores to commit their physical updates through a distributed transaction.
\Cref{algo:atomic-merge} sketches the protocol, while
\Cref{fig:atomic-merge-visibility} gives a visualization of the protocol which involves reserving
a staging interval, writing the resolved changes, and publishing them with one metadata transaction.

\parhead{Comparison and conflict resolution}
\Cref{algo:atomic-merge} begins by comparing the source and target branches against their
nearest shared fork-base branch, introduced in \Cref{sec:create}. For each key changed since
that fork base, let $B$, $S$, and $T$ denote its state at the fork base, source branch, and target
branch, respectively. \Cref{tab:merge-cases} summarizes the comparison. Conflicts are resolved
by an agent, user, or application policy before publication, producing the resolved change set
$\Delta$ that enters the staging phase in \Cref{fig:atomic-merge-visibility}(b).

\begin{table}[t]
\centering
\small
\setlength{\tabcolsep}{3pt}
\caption{Three-way Merge Cases for One Logical Key}
\label{tab:merge-cases}
\begin{tabular}{@{}ccc p{0.40\columnwidth}@{}}
\toprule
Fork base & Source branch & Target branch & Merge action \\
\midrule
$B$ & $B$ & $B$ & No action \\
$B$ & $S \ne B$ & $B$ & Apply $S$ to the target branch \\
$B$ & $B$ & $T \ne B$ & Keep $T$ \\
$B$ & $S \ne B$ & $T=S$ & Keep the common state \\
$B$ & $S \ne B$ & $T \ne B, T \ne S$ & Require conflict resolution \\
\bottomrule
\end{tabular}
\end{table}

\parhead{Coordination and reservation}
After computing $\Delta$, \Cref{algo:atomic-merge} establishes the epoch barrier from
\Cref{sec:epoch-coordination} for the source and target branches. \chronos then leverages a metadata transaction to check that neither branch has changed since the comparison and that the target branch is not
reserved by another merge. A failed check restarts the comparison; otherwise, the transaction
reserves the target branch until publication. For the example in
\Cref{fig:atomic-merge-visibility}(a), the target branch has active range
$[f_{\mathrm{old}},h)$. The transaction chooses $f_{\mathrm{new}}=f_{\mathrm{old}}+2<h$ and reserves
$I_m=[f_{\mathrm{old}}+1,h)$ for the merge. Because
$f_{\mathrm{old}}\notin I_m$, the target branch cannot yet see versions staged over this range.

\parhead{Staging and publication}
Once the target branch is reserved, \chronos writes the resolved changes durably to all
participating stores, using $I_m$ as their visibility interval
(\Cref{algo:atomic-merge} and \Cref{fig:atomic-merge-visibility}(b)). Staging applies $\Delta$ using the record-level copy-on-write mechanism as ordinary branch
updates described in \Cref{sec:writes} where it creates new physical versions, or tombstones for deletions, tagged with $I_m$,
while leaving the versions visible at $f_{\mathrm{old}}$ unchanged. If any store fails, \chronos keeps
the target branch at $f_{\mathrm{old}}$ and cancels publication, leaving all staged versions
invisible. After every store confirms durability, the final metadata transaction advances the
target branch to $f_{\mathrm{new}}\in I_m$ and clears the reservation. This is the single visibility
transition shown in \Cref{fig:atomic-merge-visibility}(c): all staged versions become visible
together, and subsequent writes use the new active range $[f_{\mathrm{new}},h)$.

\begin{algorithm}[ht]
\small
\SetKwProg{Fn}{Function}{}{}
\DontPrintSemicolon

\Fn{\textsc{merge}(source, target)}{
    $fork\_base \gets$ nearest common ancestry of source and target\;
    $preview \gets$ three-way comparison on $(fork\_base, source, target)$\;
    $resolved \gets$ resolve conflicts in $preview$ via LLM or developers\;
    establish an epoch barrier for $source$ and $target$\;
    \txnbegin{metadata transaction}\;
    validate that $preview$ is unchanged and $target$ is unreserved,
      or abort and retry\;
    $f_{\mathrm{old}} \gets frontier(target)$, $h \gets high(target)$\;
    $f_{\mathrm{new}} \gets f_{\mathrm{old}}+2$\;
    $I_m \gets [f_{\mathrm{old}}+1,h)$\tcp*[r]{staged merge versions}
    reserve $target$ and $(I_m,f_{\mathrm{new}})$ from allocator\;
    \txncommit\;
    \ForEach{$store \in$ participating stores}{
        durably stage $resolved[store]$ over $I_m$\;
    }
    on failure: clean up reservation and release epoch\;
    \txnbegin{metadata transaction}\;
    set the target branch frontier to $f_{\mathrm{new}}$ and clear its reservation\;
    \txncommit\;
    release the barrier; sessions enter the new epoch\;
}

\caption{Cross-store Merge}
\label{algo:atomic-merge}
\end{algorithm}

\parhead{Recovery and fault tolerance}
\chronos builds on the durability guarantees of the metadata store and participating data stores. We
consider crash failures and assume that a failed store eventually restarts with all writes that it
previously acknowledged as durable.

\emph{Participating-store failures.} If a participating store fails before acknowledging its staged
writes, \chronos does not publish the merge. The target branch remains at $f_{\mathrm{old}}$, and
versions staged by other stores remain hidden. The durable reservation prevents $I_m$ from being
reused while the store is unavailable. After recovery, \chronos can either finish staging and
publish after every store acknowledges durability, or remove the staged versions before clearing
the reservation. If a store fails after publication, no rollback is required: the store had already
made its staged versions durable and restores them when it restarts. Operations involving the failed
store may be unavailable during recovery, but the failure does not expose a partial merge.

\emph{Metadata-store failures.} The metadata store is the authority on whether a merge was
published. A failure before the reservation commits leaves the target branch unchanged. Once the
reservation commits, transactional recovery yields one of two states: either the target remains at
$f_{\mathrm{old}}$ with its reservation, in which case \chronos resumes or aborts the merge; or the
target has advanced to $f_{\mathrm{new}}$ and the reservation has been cleared, in which case the
merge is complete. The second outcome is safe because publication follows durable staging at every
participating store. While the metadata store is unavailable, \chronos pauses interval-changing
operations rather than risk an ambiguous allocation. Thus, failures may reduce availability but do
not expose a partially merged state under the assumptions in \Cref{sec:assumptions}.

\parhead{Correctness}
The non-overlap invariant from \Cref{eq:interval-invariant} ensures that each store exposes at most
one physical version of a logical record at either frontier. Before publication,
$f_{\mathrm{old}}\notin I_m$, so the target branch excludes every staged version. Publication occurs only
after all stores have durably prepared the versions covering $f_{\mathrm{new}}$. The epoch barrier
and the transactional frontier change then place every operation entirely before or after
publication. An operation therefore observes either the complete preceding state at
$f_{\mathrm{old}}$ or the complete merged state at $f_{\mathrm{new}}$, never a mixture of the two.

\subsection{Branch Deletion}
\label{sec:delete-gc}

Branches that are no longer useful should purge both their metadata and the physical versions that
only they can reach. Branch deletion in \chronos removes the selected branch metadata and reclaims those
versions together. Since each version is tagged with its writer identifier, \chronos can quickly locate
versions associated with the deleted branch's active ranges without scanning the entire store.

\section{IMPLEMENTATION ACROSS DATA STORES}
\label{sec:implementation}

We implement \chronos over PostgreSQL~\cite{postgres}, SQLite~\cite{sqlite},
DuckDB~\cite{duckdb}, a DBMS-backed filesystem, and Qdrant~\cite{qdrant} without modifying their
engines. Each integration maps the interval operations from \Cref{sec:algorithm,sec:branch-operations}
onto the store's native keys, filters, and atomic-update mechanisms.
We focus on the store-specific mechanisms needed for visibility filtering and interval splitting,
then briefly discuss how \chronos could extend to object storage.

\subsection{Metadata Plane and Relational Stores}
\label{sec:metadata-plane}
\label{sec:rdbms}
\label{sec:olap}

\chronos stores the branch hierarchy, interval allocations, and recovery state in transactional
tables backed by PostgreSQL or SQLite. Checkout reads this metadata to construct sessions for all
participating stores. A session then reuses the branch state throughout the branch's current epoch. After an
epoch barrier, it refreshes the frontier and active interval for all stores before admitting another
operation. In this way, ordinary data operations  do not consult the metadata plane on every access. The metadata store realizes branch epochs with a branch fence shared by sessions that
may admit operations. A branch metadata update operation proceeds immediately if it can acquire the fence; otherwise,
it asks those sessions to stop admission, waits for their current operations to drain, and then
continues. Session liveness is renewed periodically.

\parhead{Relational tables}
An application registers each logical table and its primary key, and \chronos creates a corresponding physical table containing the application's columns and the four versioning fields from \Cref{sec:model}. The application's primary key along with the lower interval bound constitutes the primary key in the physical table. \chronos then builds a secondary index over the logical key and upper bound to speed up visibility and overlap lookups. The shim for each relational store implements the query rewrites described in \Cref{sec:algorithm,sec:branch-operations} using the store's native SQL and transaction mechanisms. PostgreSQL, SQLite, and DuckDB therefore retain their native optimizers and execution paths while providing the same interval-based branch semantics.

\parhead{Schema changes}
\chronos versions the mapping from logical schemas to physical tables. A branch can modify a table
private to its current schema version directly. When the physical table is shared by multiple branches and there is a schema change request in the branch,
\chronos creates a new physical schema and copies the rows visible to that branch before applying the change.

\subsection{\chronosfs}
\label{sec:chronosfs}
Files require a different data path because applications access them through POSIX interfaces. We built \chronosfs, a file system backed by relational database~\cite{skiadopoulos2021dbos} that stores the file blocks and directory hierarchy in relational tables and exposes it through Filesystem in Userspace (FUSE)~\cite{libfuse} to offer POSIX compatibility.

\parhead{Relational representation}
\chronosfs represents inode metadata, directory entries, and files using three logical
relational tables keyed by inode identifier, parent-and-name pair, and file-and-offset pair, respectively.
These tables are then versioned similarly to the relational store implementation described in~\Cref{sec:rdbms}. \chronosfs versions file contents at 4~KiB block granularity, so writes create new versions only for
affected blocks while unchanged blocks remain shared across branches. Diff and merge use the same
internal representation but translate affected keys back into file paths and offset ranges, with text
conflicts reported as unified diffs.

\parhead{POSIX compatibility and host filesystem sandboxing}
To provide POSIX compatibility, \chronosfs implements a FUSE daemon that translates file access into
operations on the underlying relational database. Applications can therefore use ordinary POSIX interfaces
while benefiting from the kernel page cache. On a cache miss, the daemon resolves paths through the inode
and directory-entry tables and reads file contents from the block table. File-system operations that update
multiple records, such as \texttt{rename}, are implemented within a database transaction to provide atomicity.

Many agent workloads also begin with an existing directory (e.g., source code) from the host filesystem that must be made branchable. Copying the entire directory into \chronosfs would incur significant initialization overhead even though most files remain unchanged.
To avoid this cost, \chronosfs lazily copies data from host filesystem. Specifically, \chronos initially only imports the directory-tree structures without copying file contents.
Each regular file initially references its host path, and modified 4~KiB blocks are materialized as versioned records while unchanged ranges continue to reference the host file. The source directory must therefore remain read-only while any sandbox depends on it. Under this assumption, \chronosfs can branch a large host file system without duplicating its contents, while sandbox modifications remain isolated from the host directory.

\subsection{NoSQL Databases}
\label{sec:nosql-databases}
NoSQL databases, including document databases and vector databases, typically support metadata filtering. \chronos attaches interval fields to each indexed document and applies the visibility predicate before ranking. Our prototype currently supports Qdrant~\cite{qdrant}, where each document or chunk is stored as a point containing its embedding, content, and filterable payload. \chronos treats the Qdrant collection and point identifier as the logical key, stores interval metadata in the payload, and builds payload indexes for speeding up visibility and overlap lookups. When a logical point requires an update, the shim retrieves its physical versions, identifies those overlapping the active range, constructs the retained fragments and replacement in memory, and submits them through one strongly ordered Qdrant upsert. Specifically, \chronos leverages Qdrant's batch update API~\cite{qdrant-batch-update} to implement the split operation. 

\subsection{Object Stores and Data Lakes}
\label{sec:object-store}

Although we do not implement or evaluate object-store support in this work, data-lake catalogs provide a natural integration point because systems such as Iceberg and DuckLake separate logical table state from immutable object files~\cite{apache-iceberg,ducklake-specification}. A \chronos integration could version the catalog that maps tables and snapshots to physical object files, applying the standard visibility predicate to catalog queries while sharing unchanged files in S3-compatible storage~\cite{amazon-s3}. Writes would create immutable files before updating the versioned catalog, so branch creation would copy no object data and the object storage layer itself could remain unchanged. We leave this integration and its evaluation to future work.

\section{EXPERIMENTAL EVALUATION}
\label{sec:evaluation}
In this section, we conduct extensive experiments to answer the following research questions:
\begin{enumerate}[leftmargin=*,itemsep=1pt,topsep=3pt]
  \item How does \chronos compare with independently coordinated store-specific branching for cross-store agent workflows?
  \item How do branch topology and branch count affect branch creation, deletion, and query performance?
  \item What overhead does \chronos versioning impose on individual data stores?
  \item How does interval-coordinate width affect execution and storage
  efficiency?
\end{enumerate}

All experiments ran on an AWS EC2 \texttt{c5n.4xlarge} instance with 16 vCPUs, an Intel Xeon
Platinum 8124M processor, 40~GiB of memory, and Linux~6.17.

\parhead{Baselines}
We compare against five branchable relational designs: \textbf{OrpheusDB}~\cite{orpheusdb} stores immutable
records separately from version membership, requiring joins for query. \textbf{Neon} provides
copy-on-write PostgreSQL branches over shared storage~\cite{neonbranching}. \textbf{Doltgres}~\cite{doltgres} provides branching using
content-addressed storage. \textbf{PG-clone-Btrfs} creates one PostgreSQL database per branch using
Btrfs copy-on-write snapshots~\cite{btrfs-reflink}. \textbf{PG-Savepoint} emulates transient branches
with nested transactions and savepoints. For vector data, we use Branch-aware
Qdrant~\cite{qdrant-branch-aware-search} which represents branch history as part of each document's metadata and resolves it with complex filters.

\subsection{Cross-Store Enterprise Agent Workflows}
\label{sec:eval-enterprise-state}

\begin{figure}[t]
  \centering
  \begin{subfigure}[t]{0.49\columnwidth}
    \centering
    \includegraphics[width=\linewidth]{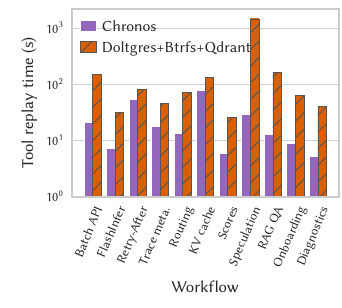}
    \par\vspace{-5pt}
    \Description{Grouped vertical bars show tool replay time on a logarithmic scale for 11
    enterprise workflows. Solid bars show Chronos and hatched bars show Doltgres plus Btrfs plus
    Qdrant.}
    \caption{Workflow replay time, excluding model inference.}
    \label{fig:enterprise-workflow-replay}
  \end{subfigure}\hfill
  \begin{subfigure}[t]{0.49\columnwidth}
    \centering
    \includegraphics[width=\linewidth]{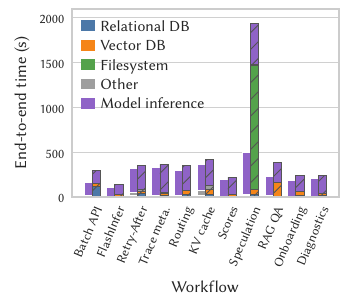}
    \par\vspace{-5pt}
    \Description{Paired stacked bars show end-to-end time for the same 11 workflows. Solid bars
    represent Chronos and hatched bars represent Doltgres plus Btrfs plus Qdrant. Each bar divides
    time among the relational database, vector database, filesystem, model inference, and other
    execution.}
    \caption{Time breakdown by component.}
    \label{fig:enterprise-workflow-breakdown}
  \end{subfigure}
  \par\vspace{-6pt}
  \caption{Enterprise agent workflows: \chronos is 5.60$\times$ faster on tool replay,
  and 1.40$\times$ faster end to end.}
  \label{fig:enterprise-workflows}
  \vspace{-0.3cm}
\end{figure}

\begin{figure}[t]
  \centering
  \begingroup
  \setlength{\fboxsep}{0pt}
  \setlength{\fboxrule}{0.4pt}
  \resizebox{\columnwidth}{!}{%
    \hbox{\fontsize{5.5}{6.5}\selectfont
      \colorbox{swarmnativeorange}{\rule{0pt}{0.6em}\hspace{1.0em}}\,
      Doltgres+Btrfs+Qdrant-Uncoordinated\quad
      \colorbox{swarmlockblue}{\rule{0pt}{0.6em}\hspace{1.0em}}\,
      Doltgres+Btrfs+Qdrant-Serialized\quad
      \colorbox{chronoslegendpurple}{\rule{0pt}{0.6em}\hspace{1.0em}}\,\chronos}}
  \endgroup
  \vspace{-0.2em}
  \begin{subfigure}[t]{0.48\columnwidth}
    \centering
    \includegraphics[width=\linewidth]{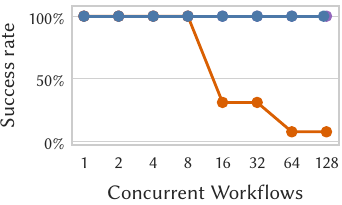}
    \par\vspace{-3pt}
    \caption{Workflow success rate}
    \label{fig:incident-response-swarm-success}
  \end{subfigure}\hfill
  \begin{subfigure}[t]{0.48\columnwidth}
    \centering
    \includegraphics[width=\linewidth]{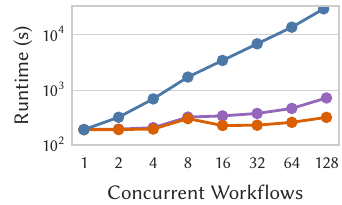}
    \par\vspace{-3pt}
    \Description{Lines show concurrent incident-investigation runtime on a logarithmic scale as concurrent
    workflows increase from one to 128 for Chronos, uncoordinated Doltgres plus Btrfs plus Qdrant,
    and serialized Doltgres plus Btrfs plus Qdrant.}
    \caption{Runtime (s)}
    \label{fig:incident-response-swarm-runtime}
  \end{subfigure}
  \par\vspace{-6pt}
  \caption{Concurrent incident investigations: at 32 workers, \chronos exposes no partial merge and
  completes in 378~s, 18.2$\times$ faster than serialization; uncoordinated execution exposes
  incomplete cross-store states.}
  \label{fig:incident-response-swarm}
  \vspace{-0.5cm}
\end{figure}

The first experiment simulates an agentic enterprise data platform whose shared data includes
company documents, emails, workplace conversations, meeting records, and source code. Agents search
this data, record durable memory, revise documents and code, and publish accepted changes to shared
organizational branches. The data spans three stores: documents and code  in a
filesystem, provenance and durable agent memory in a relational database, and document embeddings
in a vector database. We construct the company data from EnterpriseRAG-Bench~\cite{sun2026enterpriserag},
a synthetic corpus of documents and communications from an AI inference company.  We  augment it with vLLM, LiteLLM, and Langfuse source code repositories as well as their public GitHub
issues, pull requests, comments, reviews, and releases. The
resulting corpus contains 628,082 documents and 10,704,652 indexed chunks, including 97,336 GitHub
documents collected as of July~27, 2026. We organize the corpus into 23 long-lived branches for
company-wide shared data, three departments, two teams per department, and individual employees.

\begin{figure*}[t]
  \centering
  \includegraphics[width=\textwidth]{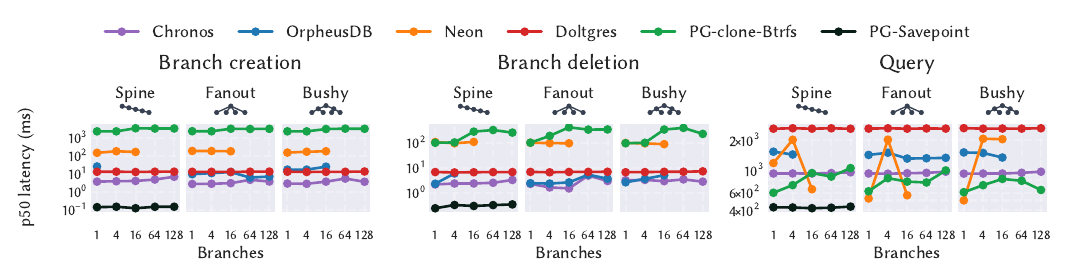}
  \vspace{-1cm}
  \Description{Branch creation, branch deletion, and query p50 latency for Chronos and five
  baselines across spine, fanout, and bushy branch topologies containing 1 to 128 branches.}
  \caption{BranchBench topology sensitivity at scale factor~5: \chronos remains stable across
  1--128 branches and tree shapes. Missing points indicate a 10-minute timeout or an unsupported shape.}
  \label{fig:branchbench-topology}
  \vspace{-0.5cm}
\end{figure*}

\begin{figure}[t]
  \centering
  \begingroup
  \scriptsize
  \setlength{\fboxsep}{0pt}
  \setlength{\fboxrule}{0.3pt}
  \begin{tabular}{@{}c@{\hspace{0.3em}}l@{\hspace{0.9em}}
                      c@{\hspace{0.3em}}l@{\hspace{0.9em}}
                      c@{\hspace{0.3em}}l@{}}
    \fcolorbox{black}{chronoslegendpurple}{\rule{0pt}{0.55em}\hspace{1.2em}} & \chronos &
    \fcolorbox{black}{chronoslegendblue}{\rule{0pt}{0.55em}\hspace{1.2em}} & OrpheusDB &
    \fcolorbox{black}{chronoslegendorange}{\rule{0pt}{0.55em}\hspace{1.2em}} & Neon \\[0.15em]
    \fcolorbox{black}{chronoslegendred}{\rule{0pt}{0.55em}\hspace{1.2em}} & Doltgres &
    \fcolorbox{black}{chronoslegendgreen}{\rule{0pt}{0.55em}\hspace{1.2em}} & PG-clone-Btrfs &
    \fcolorbox{black}{chronoslegendblack}{\rule{0pt}{0.55em}\hspace{1.2em}} & PG-Savepoint
  \end{tabular}
  \par\vspace{-2pt}
  \endgroup
  \includegraphics[width=\columnwidth]{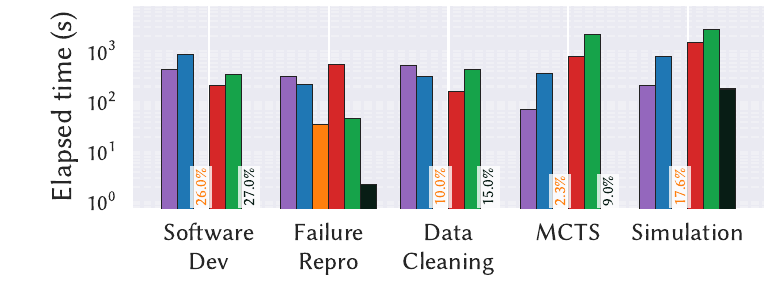}
  \par\vspace{-7pt}
  \Description{BranchBench workflow completion time for Chronos, OrpheusDB, Neon, Doltgres,
  PostgreSQL cloning on Btrfs, and PostgreSQL savepoints. Percentages report completed workflow
  steps for incomplete runs.}
  \caption{BranchBench workflows: \chronos completes all five and runs MCTS
  6.0$\times$/16.7$\times$ faster than Doltgres/PG-clone-Btrfs. Percentages report
  completed workflow steps for incomplete runs.}
  \label{fig:branchbench-macro}
  \vspace{-0.5cm}
\end{figure}

We curate 11 workflows that make concrete changes to this corpus. Eight ask an agent to resolve
coding issues in vLLM, LiteLLM, or Langfuse: the agent searches issue and company evidence, creates an
isolated branch, edits source code and tests, re-indexes the changed files, and records the reviewed
decision as durable memory. Another workflow answers company questions with cited evidence and
stores both the answers and a reusable evidence procedure. The final two create onboarding plans,
a new team charter, and employee branches from the relevant organizational branches. Across these
tasks, agents create task, team, or employee branches; update files, search indexes, and memory
records; review their changes; and then merge, delete, or retain the branches as required.

No existing baseline spans all three data stores, so we compose Doltgres for relational data,
Branch-aware Qdrant for embeddings, and Btrfs snapshots for files. In this
Doltgres+Btrfs+Qdrant baseline, the application merges data between branches independently across the three
independent stores. To isolate the effect of the branching system from model nondeterminism, we first execute each
workflow with Codex CLI~v0.145.0 using GPT-5.6-luna at maximum reasoning effort. Codex accesses the
enterprise data platform through an MCP server, and we record its model output and tool calls. We
replay the trace on each system from an identical corpus and branch hierarchy.

\parhead{Workflow performance}
\Cref{fig:enterprise-workflow-replay} shows that \chronos has significantly lower replay time for all
11 workflows, with a 5.60$\times$ geometric-mean speedup over Doltgres+Btrfs+Qdrant.
\Cref{fig:enterprise-workflow-breakdown} shows that the baseline spends 8.3$\times$ and
7.4$\times$ as much time in its relational and vector stores, respectively, because each store
maintains and resolves its own branch representation, whereas \chronos uses interval predicates
across both. Filesystem cost is more workload dependent. The largest difference occurs in
Speculation, where \chronos is 49.1$\times$ faster. This workflow compares two nested candidate
branches before selecting one to merge; deriving the two file-level differences from Btrfs snapshots
of the full corpus accounts for 94\% of the time, while \chronos restricts comparison to
versions changed by the candidates. For most workflows, model inference dominates end-to-end execution time.
Averaged across the 11 workflows, inference accounts for 91\% of \chronos's end-to-end time and
53\% of the baseline's.

\parhead{Concurrent incident investigations}
We next run concurrent  workflows that publish to one shared team branch -- these workflows simulate agents performing incident investigations. Each agent
acts as an on-call site reliability engineer for a distinct production failure. On a task, it first creates a private branch; then it searches tickets
and documents, identifies the cause, proposes a mitigation, and generates a report stored on the filesystem with metadata stored in the DBMS and an index entry in Qdrant for the report. Finally, each agent merges changes in data stores from the private branch to the shared team branch. \chronos aborts and retries transient stale-preview or reservation conflicts until it succeeds. The Uncoordinated baseline runs workflows concurrently and applies each merge independently at the three stores, which can expose partial state. The Serialized baseline instead protects each complete workflow with a global lock, preventing partial visibility at the cost of concurrency.
While benchmarking, we run a verifier that constantly checks for  partial state during merges that violate the invariant that every visible report has matching records in both the relational and vector stores. \Cref{fig:incident-response-swarm-success} reports the fraction of agent traces that
complete without violations. \chronos completes all
agents at every concurrency level through 128 agents without exposing a partially visible merge.
The uncoordinated baseline records a cross-store violation at four agents and exposes incomplete state
 from 16 agents onward; at 32 agents, it completes only 10 traces while the verifier
records 154 violations. The serialized baseline avoids these violations, but its runtime grows
linearly because workflows cannot overlap. \Cref{fig:incident-response-swarm-runtime} shows
this trade-off at 32 agents: the serialized baseline takes 6,876~s, compared with 378~s for
\chronos, an 18.2$\times$ difference. Thus, \chronos preserves cross-store visibility without
serializing complete agent executions.

\subsection{Branching Scalability and Performance}
\label{sec:eval-branching}
Next, we compare \chronos with the five relational baselines above using
BranchBench~\cite{ang2026branchbench}, which measures branch operations and workflows over TPC-C
with different branch-tree topologies.

\parhead{Topology scalability}
We start by isolating whether branch operations and query execution become more expensive as the branch tree grows.
We construct branch trees with different tree shapes. A spine extends its newest branch, a
fanout repeatedly forks from the root, and a bushy tree randomly selects parents from existing branches.
Each branch represents a what-if fulfillment plan. After creating a branch, the workload selects 100 pending orders from a random starting point, applies the corresponding delivery updates in isolation, and evaluates the resulting backlog. Branching therefore allows alternative fulfillment plans to be explored without affecting other branches.
\Cref{fig:branchbench-topology}
shows that tree shape and branch count have little effect on
\chronos. At 128 branches, \chronos creates branches 1.9--3.6$\times$ faster and deletes them
2.1--2.5$\times$ faster than Doltgres because these operations modify only compact interval metadata.
Compared with PG-clone-Btrfs, \chronos creates branches nearly three orders of magnitude faster. PG-clone-Btrfs can nevertheless
match or outperform \chronos on the queries because every clone retains PostgreSQL's native query
path. \chronos is 2.6$\times$ faster than Doltgres on the same queries at 128 branches because its
interval predicates do not depend on branch history. Thus, \chronos keeps branch operations
inexpensive without shifting a growing cost onto later updates and queries. Note that PG-Savepoint supports
only the spine shape. On completed points, OrpheusDB queries are 1.4--1.7$\times$ slower than
\chronos, and several spine and bushy runs exceed the 10-minute limit. Across Neon's completed
points through 16 branches, query latency is 1.16$\times$ that of \chronos on geometric mean, but
branch creation and deletion are 53$\times$ and 43$\times$ slower. Note that Neon only supports at most 25 branches.

\begin{figure*}[ht]
  \centering
  \begin{subfigure}[b]{0.23\textwidth}
    \centering
    \includegraphics[width=\linewidth]{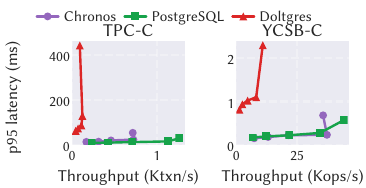}
    \caption{OLTP throughput and latency.}
    \label{fig:oltp-overhead}
  \end{subfigure}
  \hfill
  \begin{subfigure}[b]{0.177\textwidth}
    \centering
    \includegraphics[width=\linewidth]{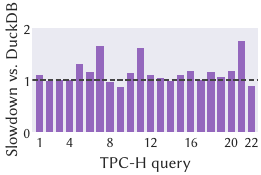}
    \caption{TPC-H.}
    \label{fig:duckdb-overhead}
  \end{subfigure}\hfill%
  \begin{subfigure}[b]{0.274\textwidth}
    \centering
    \includegraphics[width=\linewidth]{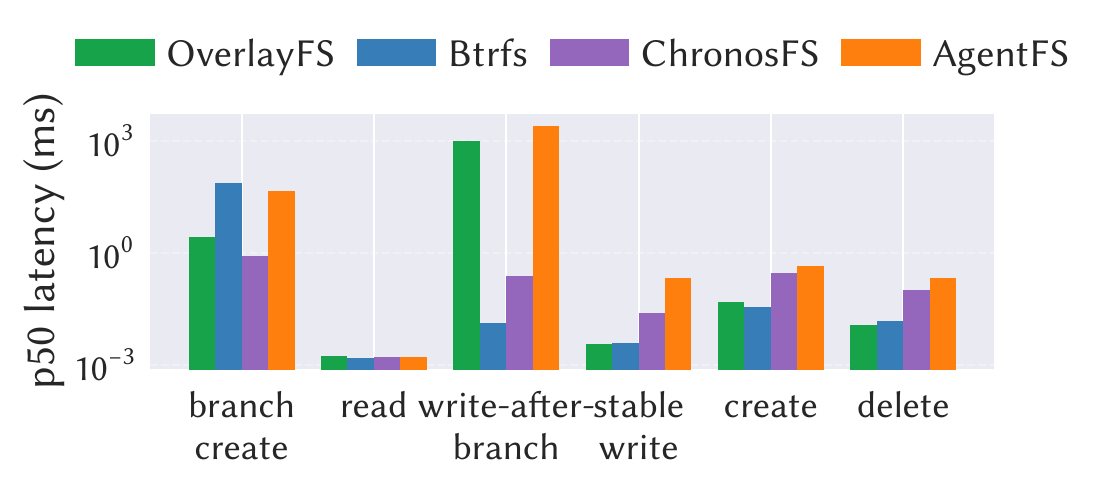}
    \caption{Filesystem p50 latency.}
    \label{fig:fs-overhead}
  \end{subfigure}\hfill%
  \begin{subfigure}[b]{0.274\textwidth}
    \centering
    \includegraphics[width=\linewidth]{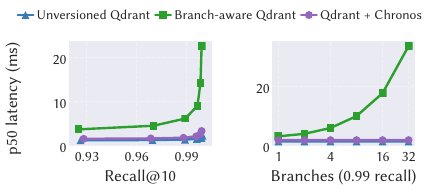}
    \caption{Qdrant latency (Dataset=DEEP).}
    \label{fig:qdrant-overhead}
  \end{subfigure}
  \par\vspace{-6pt}
  \Description{Four store-level comparisons. PostgreSQL plots p95 latency against throughput for
  TPC-C and YCSB-C. DuckDB bars show TPC-H slowdown. Filesystem bars compare operation latency.
  Qdrant plots vector-search latency against recall and branch count.}
  \caption{Store-level overhead on PostgreSQL OLTP, DuckDB TPC-H, filesystem operations, and Qdrant vector search.}
  \label{fig:operation-overhead}
  \vspace{-0.5cm}
\end{figure*}
\parhead{End-to-end workflows}
The second experiment evaluates five workflows defined in BranchBench with a 2-hour limit per system. The MCTS workflow constructs a bushy tree of 1,000 branches with
lightweight update-and-evaluate queries in each branch. The Simulation workflow creates a flat fanout of 1,000 branches.
The Software Development and Data Cleaning workflows construct moderately deep trees while applying both schema
and data transformations in each branch. The Failure Reproduction workflow creates ten short-lived branches for debugging. These
workloads therefore vary both branch intensity and the work performed within each branch. \Cref{fig:branchbench-macro} shows that \chronos completes all five workflows. PG-Savepoint
fails on three workflows, while PG-clone-Btrfs also completes all five. \chronos performs particularly well on branch-intensive workflows,
running MCTS 6.0$\times$ faster than Doltgres and 16.7$\times$ faster than PG-clone-Btrfs, and
Simulation 7.5$\times$ faster than PG-clone-Btrfs. Schema-heavy workflows remain less favorable
because \chronos copies physical tables. \chronos is 5.3$\times$ faster than OrpheusDB on MCTS workflow and 3.8$\times$ faster
on Simulation workflow. Neon completes Failure Reproduction only due to its 25-branch limit.

\subsection{Store-Level Overhead}
\label{sec:eval-overhead}
\label{sec:eval-overhead-micro}

In this subsection, we quantify the overhead added by \chronos's interval-based versioning to each data store. For OLTP, we compare \chronos with unversioned PostgreSQL and Doltgres, a PostgreSQL-compatible database with built-in branching. We also compare \chronos applied to DuckDB against unversioned DuckDB, and \chronosfs against OverlayFS, Btrfs, and AgentFS.

\parhead{PostgreSQL OLTP}
The PostgreSQL experiment uses 5 TPC-C warehouses and 5 million 1 KiB YCSB records. Each run uses a 10-second warmup followed by a 30-second measurement and reports
throughput against p95 latency. \Cref{fig:oltp-overhead} shows that \chronos incurs a
1.18--1.75$\times$ throughput slowdown over unversioned PostgreSQL while providing 3.4--5.9$\times$ the peak throughput of
Doltgres. The gap to PostgreSQL is smaller for read-only YCSB-C than for TPC-C, whose updates create
physical versions and maintain interval indexes.

\parhead{\chronos on DuckDB}
For DuckDB, we use the TPC-H benchmark with a scale factor of 10, 4 execution threads, and a 10~GiB memory limit(\Cref{fig:duckdb-overhead}). This setup reports a geometric-mean slowdown of 1.12$\times$, with per-query
ratios ranging from 0.88$\times$ to 1.76$\times$. Visibility filtering therefore adds modest
overhead on average when the data contains few physical versions, although individual queries
vary more widely.

\parhead{Qdrant vector search}
We evaluate \chronos on Qdrant using DEEP vectors~\cite{deep10M}. ~\Cref{fig:qdrant-overhead} shows the results. At 0.99 recall,
\chronos incurs a 1.34$\times$ slowdown over unversioned Qdrant and is 3.5$\times$ faster than
Branch-aware Qdrant~\cite{qdrant-branch-aware-search}. From 1 to 32 branches, \chronos latency changes
by only 1.02$\times$, while Branch-aware Qdrant grows 10.7$\times$ as its predicate size increases; at 32
branches, \chronos is 17.9$\times$ faster as its interval predicate is constant-sized.
\begin{table}[t]
    \centering
    \caption{Allocator-supported branch count across interval-coordinate widths. Counts
    exclude the root; Bushy follows BranchBench's random-parent extension.}
    \label{tab:interval-tree-capacity}
    \scriptsize
    \setlength{\tabcolsep}{2pt}
    \begin{tabular}{lrrrrrr}
      \toprule
       & \textbf{32} & \textbf{64} & \textbf{128} & \textbf{256} & \textbf{512} & \textbf{1,024} \\
      \midrule
      \textbf{Spine}  & 14 & 28 & 56 & 112 & 224 & 448 \\
      \textbf{Fanout} & $1.02{\times}10^3$ & $1.05{\times}10^6$ & $1.10{\times}10^{12}$ & $1.21{\times}10^{24}$ & $1.46{\times}10^{48}$ & $2.14{\times}10^{96}$ \\
      \textbf{Bushy}  & 61 & 498 & $1.75{\times}10^4$ & $5.49{\times}10^6$ & $>5.49{\times}10^6$ & $>5.49{\times}10^6$ \\
      \bottomrule
    \end{tabular}
    \vspace{-0.3cm}
\end{table}

\parhead{Filesystem operations}
\label{sec:eval-fs}
We compare \chronosfs backed by SQLite with OverlayFS, Btrfs, and AgentFS using sixteen 128~MiB
files. For each system, we create a branch over the files and issue one 4~KiB write to each
inherited file. We time each operation from open through close to capture its copy-on-write
cost and call this phase \emph{write-after-branch}. We then remount the branch and measure
4~KiB reads, 4~KiB writes, file creation, and file deletion. We repeat the complete experiment five times. \Cref{fig:fs-overhead} reports
the median p50 latency across runs. \chronosfs creates a branch in 0.86~ms, 3.2$\times$ faster than
OverlayFS and over 50$\times$ faster than Btrfs and AgentFS. Read latency is similar because
all four systems use the kernel page cache. Like Btrfs, \chronosfs uses block-level copy-on-write,
making write-after-branch over 4,000$\times$ faster than OverlayFS and 10,000$\times$ faster than
AgentFS, which copy the complete 128~MiB file. Relative to kernel-based Btrfs and OverlayFS, \chronosfs's overhead is mostly in
stable writes and file metadata operations.

\begin{figure}[t]
  \centering
  \begin{subfigure}[t]{0.49\columnwidth}
    \centering
    \includegraphics[width=\linewidth]{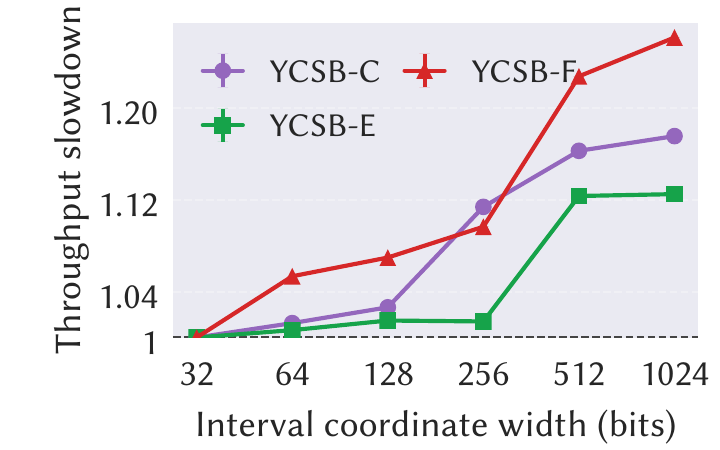}
    \caption{Throughput slowdown relative to 32-bit coordinates.}
    \label{fig:interval-width-performance}
  \end{subfigure}
  \hfill
  \begin{subfigure}[t]{0.49\columnwidth}
    \centering
    \includegraphics[width=\linewidth]{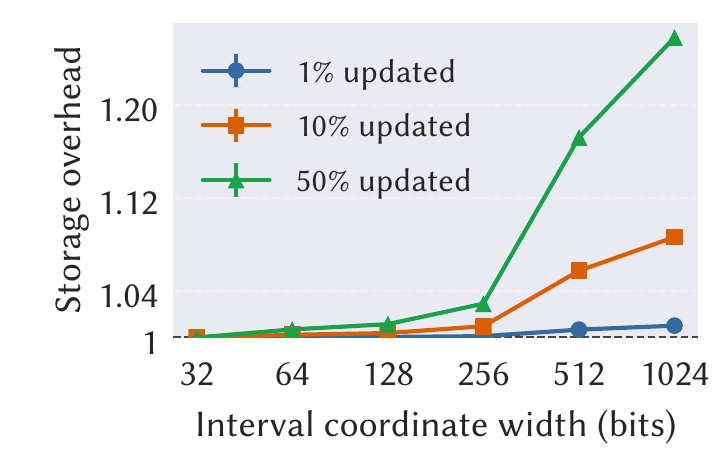}
    \caption{Storage overhead relative to 32-bit coordinates.}
    \label{fig:interval-width-storage}
  \end{subfigure}

  \Description{Two interval-coordinate width sensitivity plots. The first reports throughput
  slowdown for YCSB-C, YCSB-E, and YCSB-F. The second reports storage overhead when 1, 10, or 50
  percent of records are updated. Both compare widths from 32 to 1,024 bits.}
  \caption{Interval-coordinate width sensitivity: at 1,024 bits, throughput slowdown is at most
  1.26$\times$ and storage overhead is at most 1.26$\times$ relative to 32-bit coordinates.}
  \label{fig:interval-width}
  \vspace{-0.3cm}
\end{figure}

\subsection{Coordinate Width and Fragmentation}
\label{sec:eval-interval-width}
\parhead{Coordinate width}
We next quantify the impact of interval coordinate width with respect to query performance and storage overhead on PostgreSQL.
\chronos uses PostgreSQL's \texttt{NUMERIC} type to represent interval coordinates.
\Cref{tab:interval-tree-capacity} shows how wider interval coordinates increase the maximum
branch count supported for each tree shape.
The workload contains one million YCSB records with a 1~KiB payload and
a 16-branch spine. We use 16 workers, each operating on a separate branch and a disjoint key partition. For each coordinate width, we run YCSB-C, YCSB-E, and YCSB-F with a 10-second warmup followed by a
30-second measurement. We report throughput slowdown relative to 32-bit coordinates. To measure
storage, we update 1\%, 10\%, or 50\% of the records once across the 16 branches and report the
resulting storage overhead relative to 32-bit coordinates. \Cref{fig:interval-width-performance,fig:interval-width-storage} show that wider
coordinates impose modest overhead. At 1,024 bits, throughput slowdown ranges from 1.13$\times$ to
1.26$\times$ across workloads, while storage overhead ranges from 1.01$\times$ to 1.26$\times$
across update rates. Even with 50\% updates, storage overhead remains at most 1.03$\times$ through
256 bits. YCSB-F updates cost more because they create more interval endpoints and index
entries. PostgreSQL's variable-length \texttt{NUMERIC} type allocates storage dynamically, so its
storage consumption does not grow proportionally with the configured coordinate width.

\begin{figure}[!t]
  \centering
  \begin{subfigure}[t]{0.49\columnwidth}
    \centering
    \includegraphics[width=\linewidth]{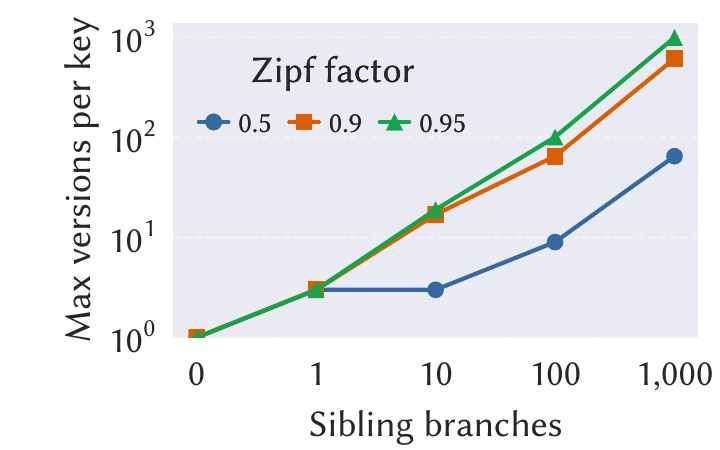}
    \caption{Live versions per key.}
    \label{fig:hot-key-fragmentation}
  \end{subfigure}
  \hfill
  \begin{subfigure}[t]{0.49\columnwidth}
    \centering
    \includegraphics[width=\linewidth]{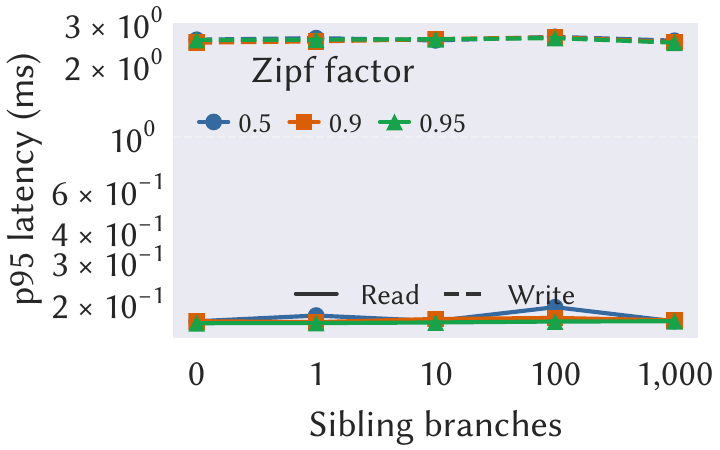}
    \caption{Read/write p95 latency.}
    \label{fig:hot-key-cost}
  \end{subfigure}
  \Description{Hot-key fragmentation under sibling updates. The left plot shows the maximum
  number of live versions for one logical key as the number of sibling branches increases under
  three Zipf factors. The right plot shows the corresponding p95 read and write latencies.}
  \caption{Hot-key fragmentation and access cost under sustained updates from sibling branches.}
  \label{fig:hot-key-amplification}
  \vspace{-0.5cm}
\end{figure}

\parhead{Hot-key fragmentation}
We next stress interval splitting by having 0--1,000 sibling branches repeatedly run YCSB-F over
the same one-million-record dataset with Zipf factors of 0.5, 0.9, and 0.95. Up to 32 branches run
concurrently, and every branch updates for 30 seconds. After compacting PostgreSQL's internal MVCC
history, we measure the live interval versions and probe YCSB-C reads and YCSB-F writes from a fresh
branch using a 10-second warmup and 30-second measurement period.
\Cref{fig:hot-key-amplification} shows that fragmentation is concentrated on the hottest keys and
increases with the number of sibling branches that modify them. At 1,000 branches, the maximum is
65, 613, and 999 live versions for Zipf factors 0.5, 0.9, and 0.95, respectively. Repeated writes
within one branch do not continually add live interval versions. Each branch maintains its own
version after the initial copy-on-write. Consequently, versions for a key are bounded by the
number of branches that modify it. Read and write
latency remains stable at 0.17ms and 2.49--2.54ms, respectively, showing that the resulting interval rows impose little additional cost at the tested scale.

\subsection{Discussion and Future Work}
\label{sec:discussion}

The results show that \chronos is most effective for applications that frequently branch and modify
records. Metadata-only branch creation and record-level sharing make these workloads substantially
faster than existing alternatives, while interval filtering keeps query overhead modest and
independent of branch history. These benefits come with two limitations. First, \chronos provides atomic cross-store visibility
under controlled access, rather than atomic
commits of physical writes. This guarantee requires controlled access through \chronos, durable
store-local writes, and a transactional metadata store. Second, \chronos's bolt-on design currently cannot enforce integrity constraints within the engines,
and semantic merge conflicts remain the application's responsibility. One possible approach is to
enforce these constraints in the \chronos layer before updates are published; we leave this to future work. Another direction for future work is a native implementation of interval-based versioning within a database. A database that only requires local branching could incorporate interval visibility
into table scans and interval splitting into its update path, eliminating shim and query-rewrite
overhead.

\section{RELATED WORK}
\label{sec:related}

\parhead{Hierarchical labeling and branched-version indexes}\allowbreak
Nested-set and nested-interval schemes use interval containment to encode
ancestry within hierarchical data~\cite{celko2004trees,tropashko2005nested}.
DeltaNI extends these schemes to versioned hierarchies~\cite{finis2013deltani}.
The BT-tree and BTR-tree use version-tree ancestry to organize records in
specialized indexes over keys, branches, and time~\cite{jiang2000bt,jiang2003btr}.
\chronos instead uses one interval representation for both branch ancestry and
record visibility. Attaching these intervals to application records reduces
visibility to a simple range predicate, enabling isolated branches across
heterogeneous stores without specialized temporal indexes.

\parhead{Database and dataset branching}
Database and dataset branching is closest to \chronos because both provide durable branches over relational data.
Doltgres and Neon provide durable branches within one relational database or storage
service~\cite{dolt,doltgres,neonbranching,ang2026branchbench}. Dataset versioning systems~\cite{datahub-cidr,datahub-pvldb,dataset-versioning-principles,decibel,orpheusdb,tardisdb} similarly provide shared versions and branches for relational data.
These systems primarily target infrequent branching within a single relational store and may incur substantial
query overhead. \chronos instead supports frequent branching across heterogeneous stores using intervals.

\parhead{Transactions and polystores}
Transactions, nested transactions, MVCC, and optimistic concurrency control provide short-lived
isolation and can emulate transient branches while an enclosing transaction remains open
~\cite{gray81,bernstein09,moss1985nested,bernstein83,kung1981optimistic}. Keeping agent candidates
open across model inference and tool execution, however, prolongs contention and delays reclamation
~\cite{kim2020long}. Polystores compose heterogeneous engines~\cite{stonebraker2015polystores,
tan2017polystore-survey,gadepally2016bigdawg,vogt2021polypheny}, while cross-store transaction
protocols coordinate atomic or serializable operations across them~\cite{georgakopoulos1991forced-conflicts,
dey2015cherry-garcia,zhang2022skeena,yamada2023scalardb,kraft2023epoxy,tang2025sonata}.
Epoxy is closest to \chronos because it also uses version metadata and read filtering across
heterogeneous stores~\cite{kraft2023epoxy}. These systems coordinate individual transactions,
whereas \chronos uses short-running transactions to manage durable branches that span many
transactions and can later be forked or merged. A merge is published through shared branch
metadata, avoiding a distributed transaction.

\parhead{Agent state and execution sandboxing}
Agent systems isolate speculative execution through copy-on-write filesystems~\cite{rodeh2013btrfs,xfs-reflink},
capability and system-call confinement~\cite{capsicum,confine}, container sandboxes~\cite{gvisor,openhands-sandbox},
microVMs~\cite{agache2020firecracker}, and agent execution environments~\cite{codex-sandbox,claude-code-sandbox,
copilot-sandbox,openhands-paper,swe-agent,e2b-sandbox}. AgentFS stores an agent's files, key-value state, and tool history in a snapshotable SQLite
database~\cite{agentfs}. These systems branch an agent's execution state, but do not coordinate
corresponding versions in external relational and NoSQL databases. \chronos complements them by
providing one durable branch across filesystem, relational, and NoSQL state.

\section{CONCLUSION}
\label{sec:conclusion}

Stateful agentic applications need to explore alternatives whose changes span heterogeneous data
stores. \chronos provides such a durable cross-store branching abstraction without modifying storage engines. \chronos introduces
interval-based versioning, which shares data and keeps query cost independent of branch history,
along with a bolt-on architecture that coordinates branch management and cross-store merge visibility.
We implement \chronos for PostgreSQL, SQLite, DuckDB, Qdrant, and a DBMS-backed filesystem.
\chronos completes every evaluated BranchBench workflow and runs MCTS 6.0$\times$ faster than
Doltgres and 16.7$\times$ faster than PostgreSQL cloning on Btrfs. These results show that efficient branching can be provided as a unified
abstraction across existing heterogeneous stores. \chronos is open-sourced at \url{https://github.com/mitdbg/chronos}.

\bibliographystyle{ACM-Reference-Format}
\bibliography{ref}

\end{document}